\documentclass[bibyear]{aa}
\usepackage[varg]{txfonts}
\usepackage{graphicx}
\usepackage{calrsfs}
\def\S{\mbox{S255IR~NIRS\,3}}

\def\Msun{\mbox{$M_\odot$}}

\def\CO{$^{12}$CO}

\def\WAT{H$_2$O}

\def\kms{\mbox{km~s$^{-1}$}}

\def\mic{\mbox{$\mu$m}}

\def\e{{\rm e}}
\def\d{{\rm d}}

\def\Tsb{T_{\rm SB}}
\def\Tb{T_{\rm B}}

\def\tho{\theta_0}
\def\thop{\theta'_0}
\def\ro{r_0}
\def\yo{y_0}
\def\ym{y_{\rm m}}
\def\ymo{y_{\rm m 0}}
\def\rmax{r_{\rm m}}
\def\rmo{r_{\rm m 0}}
\def\vo{\varv_0}

\def\no{n_0}
\def\To{T_0}
\def\xo{x_0}
\def\MM{\Lambda}
\def\Oj{\Omega\,}
\def\Mj{M}
\def\Mi{M_{\rm i}}
\def\vj{\varv}

\begin{document}

\title{
Radio outburst from a massive (proto)star
}
\subtitle{II. A portrait in space and time of the expanding radio jet from \S
\thanks{Based on observations carried out with the VLA and ALMA.}}
\author{
        R. Cesaroni\inst{1}
        \and
        L. Moscadelli\inst{1}
        \and
        A. Caratti o Garatti\inst{2,3}
        \and
        J. Eisl\"offel\inst{4}
        \and
        R. Fedriani\inst{5}
        \and
        R. Neri\inst{6}
        \and
        T. Ray\inst{3}
        \and
        A. Sanna\inst{7}
        \and
        B.~Stecklum\inst{4}
}
\institute{
 INAF, Osservatorio Astrofisico di Arcetri, Largo E. Fermi 5, I-50125 Firenze, Italy
           \email{riccardo.cesaroni@inaf.it}
\and
 INAF, Osservatorio Astronomico di Capodimonte, via Moiariello 16, I-80131 Napoli, Italy
\and
 Dublin Institute for Advanced Studies, School of Cosmic Physics, Astronomy \& Astrophysics Section, 31 Fitzwilliam Place, Dublin 2, Ireland
\and
 Th\"uringer Landessternwarte Tautenburg, Sternwarte 5, D-07778 Tautenburg, Germany
\and
 Instituto de Astrof\'{\i}sica de Andaluc\'{\i}a, CSIC, Glorieta de la Astronom\'{\i}a s/n, E-18008 Granada, Spain
\and
 Institut de Radioastronomie Millim\'etrique (IRAM), 300 rue de la Piscine, F-38406 Saint Martin d’H\`eres, France
\and
 INAF, Osservatorio Astronomico di Cagliari, Via della Scienza 5, I-09047 Selargius (CA), Italy
}
\offprints{R. Cesaroni, \email{riccardo.cesaroni@inaf.it}}
\date{Received date; accepted date}

\abstract{
Growing observational evidence indicates that the accretion process
leading to star formation may occur in an episodic way, through
accretion outbursts revealed in various tracers. This phenomenon
has also now been detected in association with a few young massive (proto)stars  ($>$8~\Msun), where an increase in the emission has been observed from
the IR to the centimetre domain. In particular, the recent outburst at
radio wavelengths of \S\ has been interpreted as due to the expansion of a
thermal jet, fed by part of the infalling material, a fraction of which
has been converted into an outflow.
}{
We wish to follow up on our previous study of the centimetre and millimetre
continuum emission from the outbursting massive (proto)star \S\ and
confirm our interpretation of the radio outburst, based on an expanding
thermal jet.
}{
The source was monitored for more than 1~yr in six bands from 1.5~GHz to
45.5~GHz with the \textit{Karl G. Jansky} Very Large Array, and, after an interval of
$\sim$1.5~yr, it was imaged with the Atacama Large Millimeter/submillimeter
Array at two epochs, which made it possible to detect the proper motions of
the jet lobes.
}{
The prediction of our previous study is confirmed by the new results.
The radio jet is found to expand, while the flux, after an initial exponential
increase, appears to stabilise and eventually decline, albeit very slowly.
The radio flux measured during our monitoring is
attributed to a single lobe, expanding towards the NE. However, starting from
2019, a second lobe has been emerging in the opposite direction, probably powered
by the same accretion outburst as the NE lobe, although with a delay of at
least a couple of years. Flux densities measured at frequencies higher than
6~GHz were satisfactorily
fitted with a jet model, whereas those below 6~GHz are clearly
underestimated by the model. This indicates that non-thermal
emission becomes dominant at long wavelengths.
}{
Our results suggest that thermal jets can be a direct consequence of
accretion events, when yearly flux variations are detected. The formation of
a jet lobe and its early expansion appear to have been triggered by the accretion
event that started in 2015. The end of the accretion outburst is also mirrored
in the radio jet. In fact, $\sim$1~yr after the onset of the radio outburst,
the inner radius of the jet began to increase, at the same time the
jet mass stopped growing, as expected if the powering mechanism of the
jet is quenched.  We conclude that our findings strongly support a tight
connection between accretion and ejection in massive stars, consistent with
a formation process involving a disk--jet system similar to that of low-mass stars.

}

\maketitle

\section{Introduction}
\label{sint}

In recent years, observations have provided us with increasing evidence of
circumstellar rotating structures around B-type (proto)stars, especially
since the advent of the Atacama Large Millimeter/submillimeter
Array (ALMA; see e.g.
the review by Beltr\'an \& de Wit~\cite{beldew}). This strongly
suggests that disk-mediated accretion could be a viable mechanism to feed
even the most massive stars. However, we still do not understand the physical properties of these rotating
structures, nor how such an accretion proceeds, whether through a smooth, continuous flow or
episodically with parcels of material falling onto the star. The
recent detection of outbursts (Stecklum~\cite{steck16,steck21}; Caratti
o Garatti et al.~\cite{cagana}; Hunter et al.~\cite{hunt17,hunt21};
Burns et al.~\cite{burns20,burns23}; Chen et al.~\cite{chen21}) in a few
luminous young stellar objects (YSOs) $>$8~\Msun\ provides us with the
intriguing possibility that these phenomena could be the consequence of
episodic accretion events, akin to those commonly observed in low-mass
YSOs as FU~Orionis and EX~Orionis events (Audard et al.~\cite{aud}; Fischer et
al.~\cite{fischppvii}).

In this study we focus on the outburst from the massive (proto)star
\S, located at a distance of $1.78^{+0.12}_{-0.11}$~kpc (Burns et
al.~\cite{burns16}). This object is unique because the outburst has
been observed not only in the emission of some maser species (Fujisawa et
al.~\cite{fuji}; Moscadelli et al.~\cite{mosca17}; Hirota et al.~\cite{hiro21})
and in lines and continua at IR
and (sub-)millimetre wavelengths (Caratti~o~Garatti et al.~\cite{cagana};
Uchiyama et al.~\cite{uchi}; Liu et al.~\cite{liu18}), but also in the
centimetre domain (Cesaroni et al.~\cite{cesa18}; hereafter Paper~I). A
time delay of $\sim$1~yr was found between the onset of the IR and radio
outbursts, consistent with the different mechanisms at the origin of the two:
in fact, the IR outburst is based on radiative processes propagating at velocities
comparable to the speed of light, whereas the radio
outburst is due
to shocks expanding approximately at the speed
of sound. In Paper~I we present compelling evidence of an exponential
increase in the radio emission from the thermal jet associated with
this source. Given the existence of a disk-jet system in \S\ (Boley et
al.~\cite{bole13}; Wang et al.~\cite{wang}; Zinchenko et al.~\cite{zin15};
Liu et al.~\cite{liu20}), we believe that we are witnessing an episodic
accretion event mediated by the disk, where part of the infalling material
has been diverted into the associated jet (Fedriani et al.~\cite{fedr23}). In
Paper~I
we show that a simple model of an expanding jet can
satisfactorily reproduce the increase in the radio flux observed in four
bands. Although the emission was basically unresolved, we predicted that
in a few years the jet expansion should make it possible to resolve its
structure. With this in mind, we performed both monitoring of the radio
emission and sub-arcsecond millimetre imaging at two epochs, several
years after the beginning of the outburst. In this article we report on
the results of these observations.

\section{Observations}
\label{sobs}

The radio emission was monitored with the \textit{Karl G. Jansky} Very Large Array
(VLA) and, at a later time, with ALMA.
In the following we describe the observational setup and
data reduction separately for the two datasets.
In both cases, for the phase centre we chose the position
$\alpha$(J2000)=$06^{\rm h}\,12^{\rm m}\,54\fs02$,
$\delta$(J2000)=17\degr\,59\arcmin\,23\farcs1.

\subsection{Very Large Array}
\label{svla}

\S\ was observed at 11 epochs from January 2017 to January 2018
(project codes: 16B-427 and 17B-045). The observing dates and array
configurations are listed in Table~\ref{tflux}, where for the sake of
completeness we have also included the 2016 data already presented in
Paper~I. We note that the table contains as yet unpublished data in the L band
obtained in the same observing run (project 16A-424) described in Paper~I.

The signal was recorded with the Wideband Interferometric Digital
ARchitecture (WIDAR) correlator in six bands, centred approximately at 1.5 (L
band), 3 (S), 6 (C), 10 (X), 22.2 (K), and 45.5~GHz (Q), in dual polarisation
mode. The total observing bandwidth (per polarisation) was 1~GHz in the L band,
2~GHz in the S band, 4~GHz in the C and X bands, and 8~GHz in the K and Q
bands.
The
primary flux calibrator was 3C48 and the phase-calibrators were J0632+1022
in the L and S bands, J0559$+$2353 in the C and X bands, and J0539+1433 in the K and
Q bands.

We made use of the calibrated dataset provided by the NRAO pipeline
and subsequent inspection of the data and imaging were performed with the
CASA\footnote{The Common Astronomy Software Applications software can be
downloaded at http://casa.nrao.edu} package,
version 5.6.2-2.
The continuum images were
constructed using natural weighting to maximise flux recovery. Typical
values of the 1$\sigma$ RMS noise level and synthesised beam are given in
Table~\ref{tnoise} for each band and array configuration.

\begin{table*}
\caption[]{
Flux densities of \S\ observed at different epochs and frequencies.
}
\label{tflux}

\begin{tabular}{cccccccc}
\hline
\hline
year: & & \multicolumn{6}{c}{2016$^a$} \\
\cline{1-1} \cline{3-8}
date: & & Mar 11 & Jul 10 & Aug 1 & Oct 15 & Nov 20 & Dec 27 \\
days: & & --121 & 0 & 22 & 97 & 133 & 170 \\
array: & & C & B & B & A & A & A \\
\cline{1-1} \cline{3-8}
$\nu$ (GHz) & & \multicolumn{6}{c}{$S_\nu$ (mJy)} \\
\cline{1-1} \cline{3-8}
1.5 & & ---  & ---  & --- & 3.58 &  3.3 &  4.7 \\
3.0 & & ---  & ---  & --- & ---  & ---  & ---  \\
6.0 & & 0.87 & 0.79 & 1.1 & 2.63 &  4.8 &  7.9 \\
10.0 & & 1.0 & 1.2  & 1.8 & 3.49 &  6.7 & 11.0 \\
22.2 & & 2.7 & 2.2  & 3.1 & 6.58 & 11.3 & 17.6 \\
45.5 & & 4.4 & 1.9  & 3.8 & 13.6 & 21.3 & 26.8 \\
\hline
\end{tabular}

\vspace*{1mm}
$^a$ Data acquired in 2016 at $\nu\ge6$~GHz have already been reported
by Cesaroni et al.~(\cite{cesa18}) and are listed here for the sake of
completeness. \\

\vspace*{3mm}

\begin{tabular}{cccccccccccc}
\hline
\hline
year: & & \multicolumn{10}{c}{2017} \\
\cline{1-1} \cline{3-12}
date: & & Jan 26 & Mar 3 & Mar 19 & Jul 29 & Aug 19 & Aug 25 & Sep 17 & Oct 23 & Dec 8 & Dec 27 \\
days: & & 200 & 236 & 252 & 384 & 405 & 411 & 434 & 470 & 516 & 535 \\
array: & & A & D & D & C & C & C & B & B & B & B \\
\cline{1-1} \cline{3-12}
$\nu$ (GHz) & & \multicolumn{10}{c}{$S_\nu$ (mJy)$^b$} \\
\cline{1-1} \cline{3-12}
 1.5 & &  3.9 & $<$8.5 & $<$14.9 & $<$14.2 & $<$17.4 & $<$10.8 & $<$7.5 & $<$9.3 & $<$8.7 & $<$7.3 \\
 3.0 & &  6.1 & $<$9.9 & $<$13.8 & $<$11.2 & $<$14.1 & $<$13.4 & 14.6 & 14.7 & 14.5 & 13.4 \\
 6.0 & &  8.1 &  9.2$^c$ &  10.0$^c$ &  14.3 &  16.3 &  16.1 & 17.7 & 18.5 & 16.9 & 17.4 \\
10.0 & & 11.4 & 12.8$^d$ &  12.8$^d$ &  20.9 &  22.3 &  22.6 & 24.0 & 23.1 & 22.7 & 22.7 \\
22.2 & & 20.4 & 20.7 &  22.8 &  30.0 &  34.0 &  36.8 & 29.5 & 29.5 & 27.6 & 26.9 \\
45.5 & & 26.8 & 28.9 &  34.5 &  35.9 &  44.1 &  45.3 & 26.0 & 28.3 & 24.2 & 23.3 \\
\hline
\end{tabular}

\vspace*{1mm}
$^b$ The symbol `$<$' indicates that the value also includes the
contribution from the two large-scale lobes, due to insufficient angular
resolution. \\
$^c$ Corrected by subtracting a fiducial value of 1.8~mJy for the two large-scale lobes (see main text). \\
$^d$ Corrected by subtracting a fiducial value of 1.0~mJy for the two large-scale lobes (see main text).\\

\vspace*{3mm}

\begin{tabular}{ccccccccccc}
\hline
\hline
year: & & \multicolumn{3}{c}{2018} & & 2019 & & 2021 \\
\cline{1-1} \cline{3-5} \cline{7-7} \cline{9-9}
date: &  & Jan 15 & Mar 29 & May 5 & & Jun 6 & & Sep 3 \\
days: &  & 554 & 627 & 664 & & 1061 & & 1881 \\
array: & & B & A & A & & C43-9/10 & & C43-9/10 \\
\cline{1-1} \cline{3-5} \cline{7-7} \cline{9-9}
$\nu$ (GHz) & & \multicolumn{3}{c}{$S_\nu$ (mJy)} & & $S_\nu$ (mJy) & & $S_\nu$ (mJy) \\
\cline{1-1} \cline{3-5} \cline{7-7} \cline{9-9}
1.5 &  & $<$6.9$^b$ & ---  & ---  & & ---  & & ---  \\
3.0 &  & 13.3 & ---  & ---  & & ---  & & ---  \\
6.0 &  & 16.8 & 15.1$^e$ & ---  & & ---  & & ---  \\
10.0 & & 22.2 & ---  & ---  & & ---  & & ---  \\
22.2 & & 25.0 & ---  & 24.7$^e$ & & ---  & & ---  \\
45.5 & & 23.3 & ---  & ---  & & ---  & & ---  \\
92.2 & & ---  & ---  & ---  & & 46.0 (22.8)$^f$ & & 42.3 (18.8)$^f$ \\
\hline
\end{tabular}

\vspace*{1mm}
$^e$ From Obonyo et al.~(\cite{obonyo}). \\
$^f$ The value in parentheses is the flux density of the NE lobe only (see Sect.~\ref{smmem}). \\[2mm]
Note: The row labelled with `days' gives the number of days after the
beginning of the radio burst. The error on the flux densities is estimated
to be $\sim$10\% (see main text).

\end{table*}

The flux density of the compact, variable source centred on \S\ has been
estimated inside a polygonal shape encompassing the compact, unresolved
radio source and is given in Table~\ref{tflux} for each band and epoch.
As explained in Paper~I, in our VLA observations the continuum emission
from \S\ appears as an unresolved component (the variable source of interest
for our study) plus two large-scale lobes separated by $\sim$8\arcsec,
or $\sim$15000~au, in projection (see Fig.~1 of Paper~I) that in all
likelihood are originating from a previous outburst that occurred many decades
ago. At the longest wavelengths it was possible to resolve the central
source from the lobes only in the most extended array configuration. In
the other cases, we could only measure the total flux density from all
components and the corresponding values are reported in Table~\ref{tflux}
as upper limits. In a few cases, as indicated in the footnotes of the
table, we attempted a correction to the flux, under the assumption that
the flux of the lobes is constant in time and thus equal to the value measured
(at a different epoch) with the A array. With this approach a caveat is in
order, because a compact configuration may be sensitive to lobe structures
that are resolved out in a more extended configuration. This implies that
the corrected flux densities could be still an upper limit.

In order to estimate the uncertainty on the flux density measurements
of \S, we took advantage of the presence of a compact, marginally resolved
continuum source located approximately at
$\alpha$(J2000)=$06^{\rm h}\,12^{\rm m}\,53\fs61$,
$\delta$(J2000)=18\degr\,00\arcmin\,26\farcs4,
which happens to fall in the primary beam in all bands except
band Q. Under the reasonable assumption that this object is not
variable, by comparing the flux density measurements
obtained at different epochs at the same frequency, we estimated
a relative error of 10\% in all bands.

\subsection{Atacama Large Millimeter/submillimeter
Array}
\label{salma}

\S\ was observed with ALMA in band~3 on June 6, 2019, and September 3, 2021
(project 2018.1.00864.S, P.I. R.~Cesaroni). The main characteristics of
the observations are summarised in Table~\ref{talma}. The correlator was
configured with 4 units of 2~GHz, in double polarisation, centred at 85.2,
87.2, 97.2, and 99.2~GHz. The spectral resolution is 0.49~MHz (corresponding
to $\sim$1.5--1.7~\kms, depending on the frequency), sufficient to identify
line-free channels and obtain a measurement of the continuum emission.

The data were calibrated through the ALMA data reduction
pipeline. For each 2~GHz correlator unit, we created a data cube using
task {\it tclean} of CASA, adopting natural weighting and a circular
beam of 0\farcs097 for the first epoch and 0\farcs087 for the second one.
To create a continuum map for each 2~GHz band, we used the STATCONT
software\footnote{https://hera.ph1.uni-koeln.de/$\sim$sanchez/statcont}
developed by S\'anchez-Monge et al.~(\cite{statcont}). In this way
we also obtained cubes of the continuum-subtracted line emission. Finally,
the four continuum maps were averaged together to increase the
signal-to-noise ratio.  The measured fluxes at the two epochs are reported
in Table~\ref{tflux}.

We note that the derivation of the continuum images described above also provides
us with continuum-subtracted channel maps. Although a study of the
line emission in this region goes beyond the purposes of this article, in
the following we briefly consider the maps of two molecular transitions,
SO(2$_2$--1$_1$) at 86093.983~MHz and H$^{13}$CN(1--0) at 86338.7~MHz.

\begin{table}
\centering
\caption[]{
Typical 1$\sigma$ RMS noise and synthesised beam for different
bands and VLA configurations.
}
\label{tnoise}
\begin{tabular}{ccccc}
\hline
\hline
band -- $\nu$(GHz) & \multicolumn{4}{c}{noise, beam} \\
                   & \multicolumn{4}{c}{($\mu$Jy/beam), (arcsec)} \\
\cline{2-5}
 & A & B & C & D \\
\hline
 L -- 1.5 & 84, 1.4 & 53, 4.7 & 148, 15 & 662, 44 \\
 S -- 3 & 27, 0.69 & 31, 2.3 &  68, 7.2 & 347, 23 \\
 C -- 6 & 11, 0.37 & 24, 1.2 &  67, 3.8 & 693, 12 \\
 X -- 10 &  8, 0.21 & 14, 0.71 &  37, 2.3 & 342, 7.7 \\
 K -- 22.2 & 16, 0.094 & 19, 0.36 &  36, 1.1 & 111, 3.4 \\
 Q -- 45.5 & 81, 0.061 & 42, 0.20 &  51, 0.57 & 102, 1.9 \\
\hline
\end{tabular}
\end{table}

\begin{table}
\centering
\caption[]{
Main parameters of the ALMA observations.
}
\label{talma}
\begin{tabular}{ccc}
\hline
\hline
date & 2019/6/6 & 2021/9/3 \\
\hline
 configuration & C43-9/10 & C43-9/10 \\
 antennae & 43 & 47 \\
 baselines (m) & 237--1524 & 122--1619 \\
 flux cal. & J0750+1231 & J0750+1231 \\ 
 bandpass cal. & J0750+1231 & J0750+1231 \\
 phase cal. & J0613+1708 & J0613+1708 \\
 synthesised beam & 0\farcs097 & 0\farcs087 \\
 1$\sigma$ RMS noise & 50~$\mu$Jy/beam & 30~$\mu$Jy/beam \\
\hline
\end{tabular}
\end{table}
 
\section{Results}
\label{sres}

\subsection{Continuum emission at $\lambda=3$~mm}
\label{srmm}

\begin{figure}
\centering
\resizebox{8.5cm}{!}{\includegraphics[angle=0]{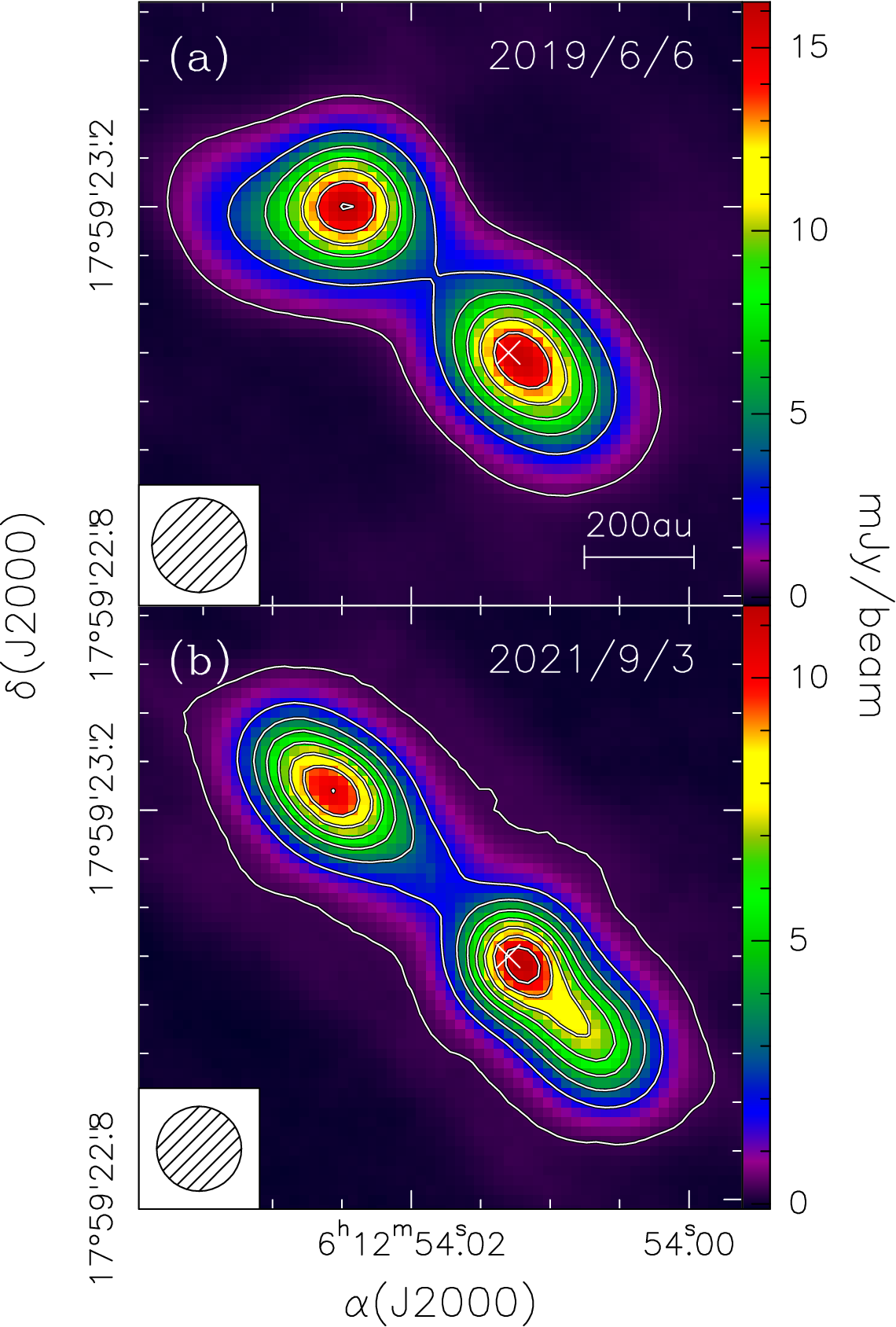}}
\caption{
Maps of the 3~mm continuum emission obtained with ALMA.
{\bf a.} Data acquired on June 6, 2019.
Contour levels range from 0.52 to 16.12 in steps of 2.6~mJy/beam.
The cross marks the peak of the 900~\mic\ continuum emission
imaged by Liu et al.~(\cite{liu20}).
The circle in the bottom left represents the synthesised beam.
{\bf b.} Same as the top panel, but for the map obtained on September 3, 2021.
Contour levels range from 0.29 to 16.24 in steps of 1.45~mJy/beam.
}
\label{falma}
\end{figure}

\begin{figure}
\centering
\resizebox{8.5cm}{!}{\includegraphics[angle=0]{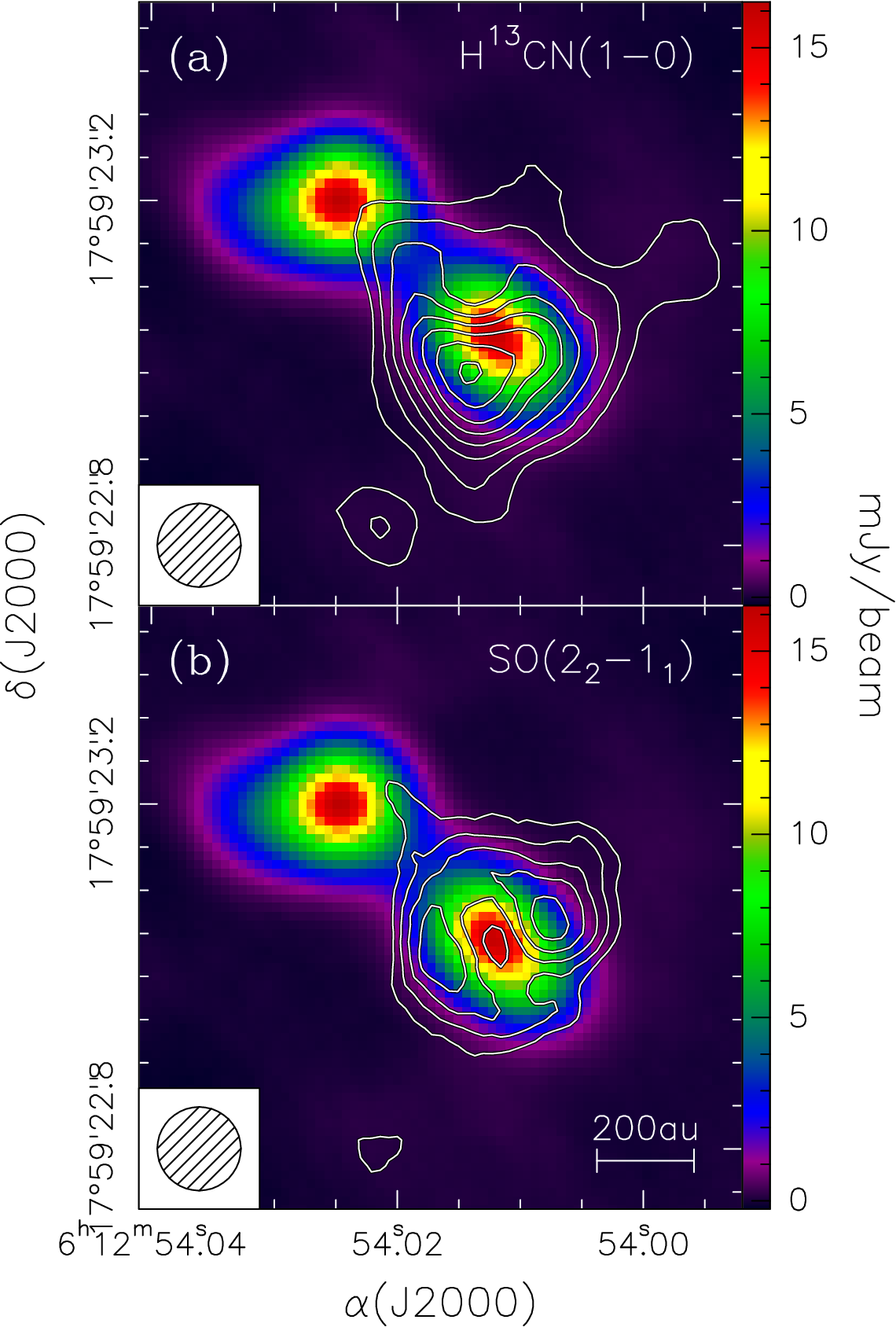}}
\caption{
Maps of the line emission obtained with ALMA.
{\bf a.} Map of the emission averaged over the H$^{13}$CN(1--0) line (white
contours) overlaid on
the continuum image at 3~mm obtained on June 6, 2019.
Contour levels range from 1.2 to 4.56 in steps of 0.48~mJy/beam.
The circle in the bottom left represents the synthesised beam.
{\bf b.} Same as the top panel, but for the SO(2$_2$--1$_1$) line.
Contour levels range from 1.65 to 4.62 in steps of 0.33~mJy/beam.
}
\label{fmols}
\end{figure}

The structure of the radio jet from \S\ at the two epochs observed with
ALMA is shown in Fig.~\ref{falma}. The most striking result is the clear
expansion of the jet, which becomes significantly more elongated to both the
NE and SW. At first look, the figure seems to outline a bipolar structure,
where the star might be located close to the geometrical centre,
between the two jet lobes. However, this is not the case. Careful comparison
between Fig.~\ref{falma}a and~\ref{falma}b reveals that, while the NE peak
is moving away from the centre, the SW peak stays still, consistent with
this being the location of the (proto)star. This result is confirmed by
the coincidence of the SW peak with the peak of the sub-millimetre emission (white
cross) measured by Liu et al.~(\cite{liu20}), as expected if the (proto)star
lies inside the parental molecular, dusty core. Figure~\ref{fmols} clearly
confirms this scenario by showing an overlay of the molecular emission maps
of the SO(2$_2$--1$_1$) and H$^{13}$CN(1--0) lines observed by us, with
an image of the 3~mm continuum emission. It is worth noting that in both
transitions a dip is seen towards the continuum peak,
which suggests that part of the line emission is likely absorbed against the
bright continuum. This is consistent with the high brightness temperature
of both 3~mm continuum peaks, of the order of 500~K (obtained from the
maps cleaned with uniform weighting).

With all the above in mind, in the following we assume that the
(proto)star powering the radio jet is located at the SW peak of the
3~mm continuum emission, namely at
$\alpha$(J2000)=$06^{\rm h}\,12^{\rm m}\,54\fs012$,
$\delta$(J2000)=17\degr\,59\arcmin\,23\farcs04. We prefer to use the
peak position of our maps instead of that of the sub-millimetre maps of Liu et
al.~(\cite{liu20}), because of the higher angular resolution (0\farcs087
instead of 0\farcs14).

\begin{table}
\centering
\caption[]{
Position angles of the jet outflow from \S\ in different tracers.
}
\label{tpa}
\begin{tabular}{cccc}
\hline
\hline
tracer & size & PA & ref. \\
\hline
 \CO(2--1) & 40\arcsec (0.35 pc) & 75\degr & Wang et al.~(\cite{wang}) \\
 cm cont. & 8\arcsec (0.07 pc) & 70\degr & Paper~I \\
 mm cont. & 0\farcs6 (1000 au) & 48\degr & this paper \\
 \WAT\ masers & 0\farcs5 (900 au) & 44\degr & Hirota et al.~(\cite{hiro21}) \\
\hline
\end{tabular}
\begin{flushleft}
\end{flushleft}
\end{table}
 
\begin{figure} 
\centering
\resizebox{8.5cm}{!}{\includegraphics[angle=0]{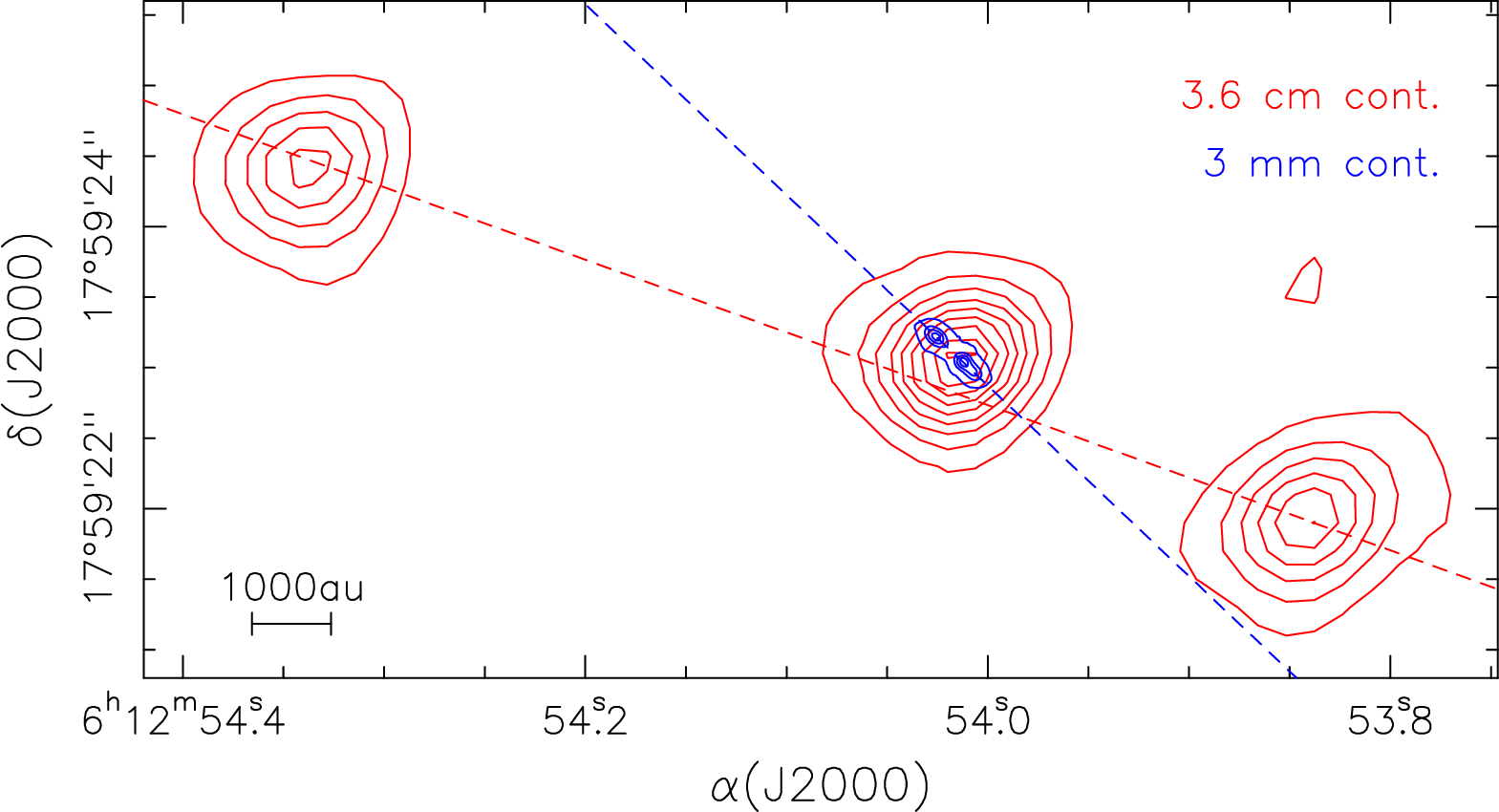}}
\caption{
Maps of the 3.6~cm continuum emission obtained with the VLA
on July 10, 2016 (red) and of the 3~mm continuum emission
(blue), also shown in Fig.~\ref{falma}b.
The dashed lines denote the symmetry axes of the bipolar structures.
Red contour levels range from 0.1 to 0.6 in steps of 0.1~mJy/beam.
Blue contour levels range from 0.29 to 8.99 in steps of 2.9~mJy/beam.
}
\label{fjets}
\end{figure}

It is also interesting to compare the jet lobes observed in our ALMA
maps with those seen on a much larger scale (see Fig.~1 of Paper~I). The
comparison is presented in Fig.~\ref{fjets}, where we show the maps of
the 3.6~cm and 3~mm continuum emission. Clearly, the directions of the
symmetry axes of the two pairs of lobes are remarkably different, with
position angle (PA) of $\sim$70\degr, on the large scale, and $\sim$48\degr,
on the small scale. The fact that the two axes have different orientations
and do not intersect at the
position of the star (i.e. at the SW peak of the 3~mm continuum) can be
interpreted in two ways: either we are dealing with two jets originating
from two different YSOs, or the jet is precessing and the star is moving on
the plane of the sky at a different speed with respect to the large-scale
lobes. The latter scenario implies that either the star or the ejected
material is experiencing deceleration or acceleration, such that one of the
two is lagging behind the other.

\begin{figure*}[h]
\centering
\resizebox{\hsize}{!}{\includegraphics[angle=0]{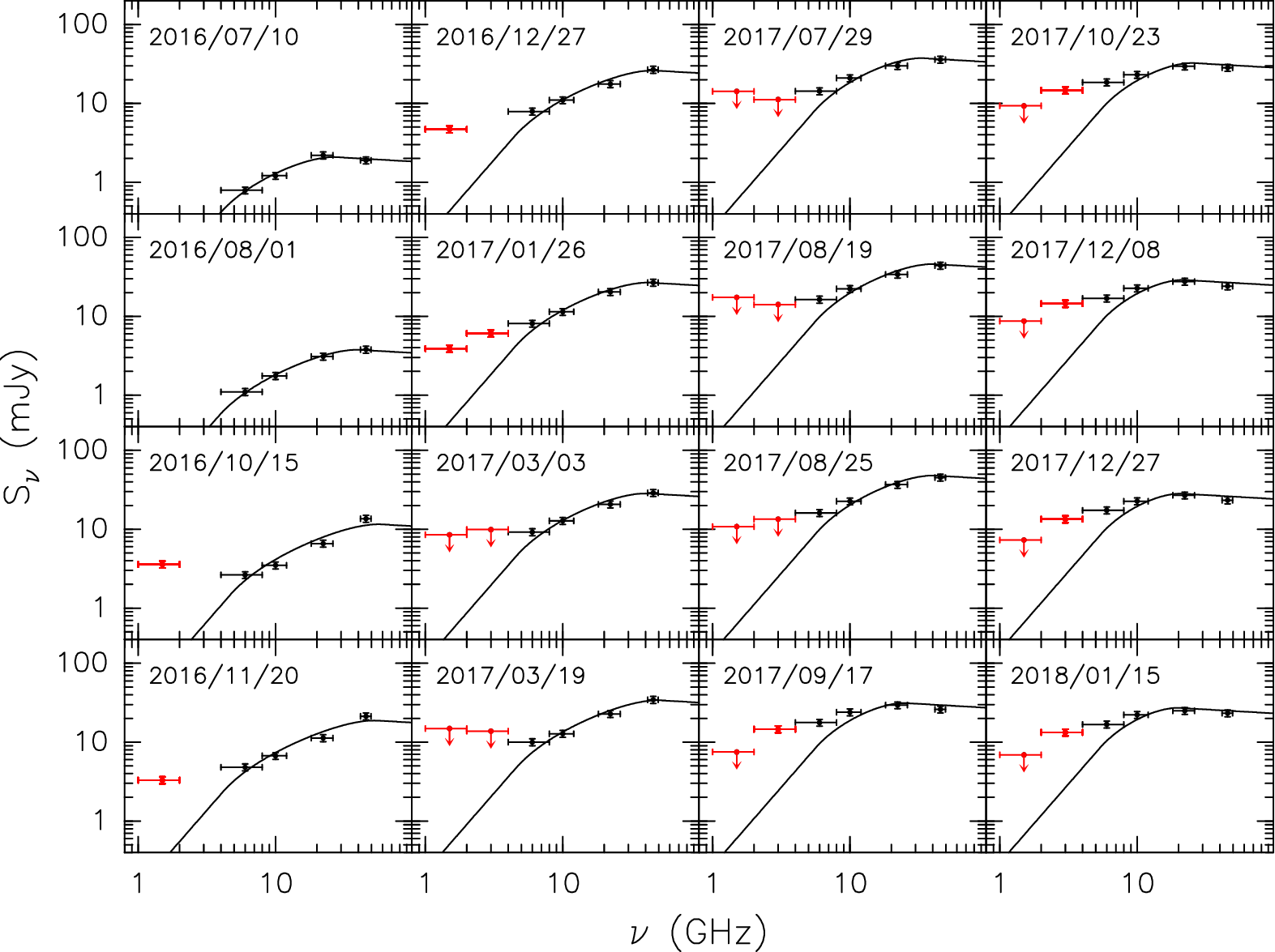}}
\caption{
Continuum spectra of the variable, compact component in \S. The date of
the observations is given in the top left of each panel. The vertical error
bars assume a calibration error of 10\% in all bands, while the horizontal
bars indicate the bandwidth covered at each frequency. The upper limits
denote that the flux density is also contributed by the emission from the
large-scale lobes, due to the limited angular resolution. Red symbols
mark data in the L (1.5~GHz) and S (3~GHz) bands. The curves are the best
fits obtained with the model of a thermal jet described in Sect.~\ref{smod}.
}
\label{fsedfits}
\end{figure*}

Although it is impossible to rule out one of the two hypotheses, in the following we assume that all lobes belong to the same jet, because there is only one
core lying along the jet axis and evidence for precession has been found
in \S\ (Wang et al.~\cite{wang}; Fedriani et al.~\cite{fedr23}), as well as
other similar objects (see e.g. Shepherd et al.~\cite{shep00},
Cesaroni et al.~\cite{cesa05}, S\'anchez-Monge et al.~\cite{sanch14},
Beltr\'an et al.~\cite{bel16}). Further support for this hypothesis is given
by the progressive change of the position angle of the jet from
the large to the small scale, as shown in Table~\ref{tpa}. This trend is
indeed consistent with a jet outflow undergoing precession.

\subsection{Continuum emission at $\lambda\ge7$~mm}
\label{srcm}

\begin{figure}
\centering
\resizebox{8.5cm}{!}{\includegraphics[angle=0]{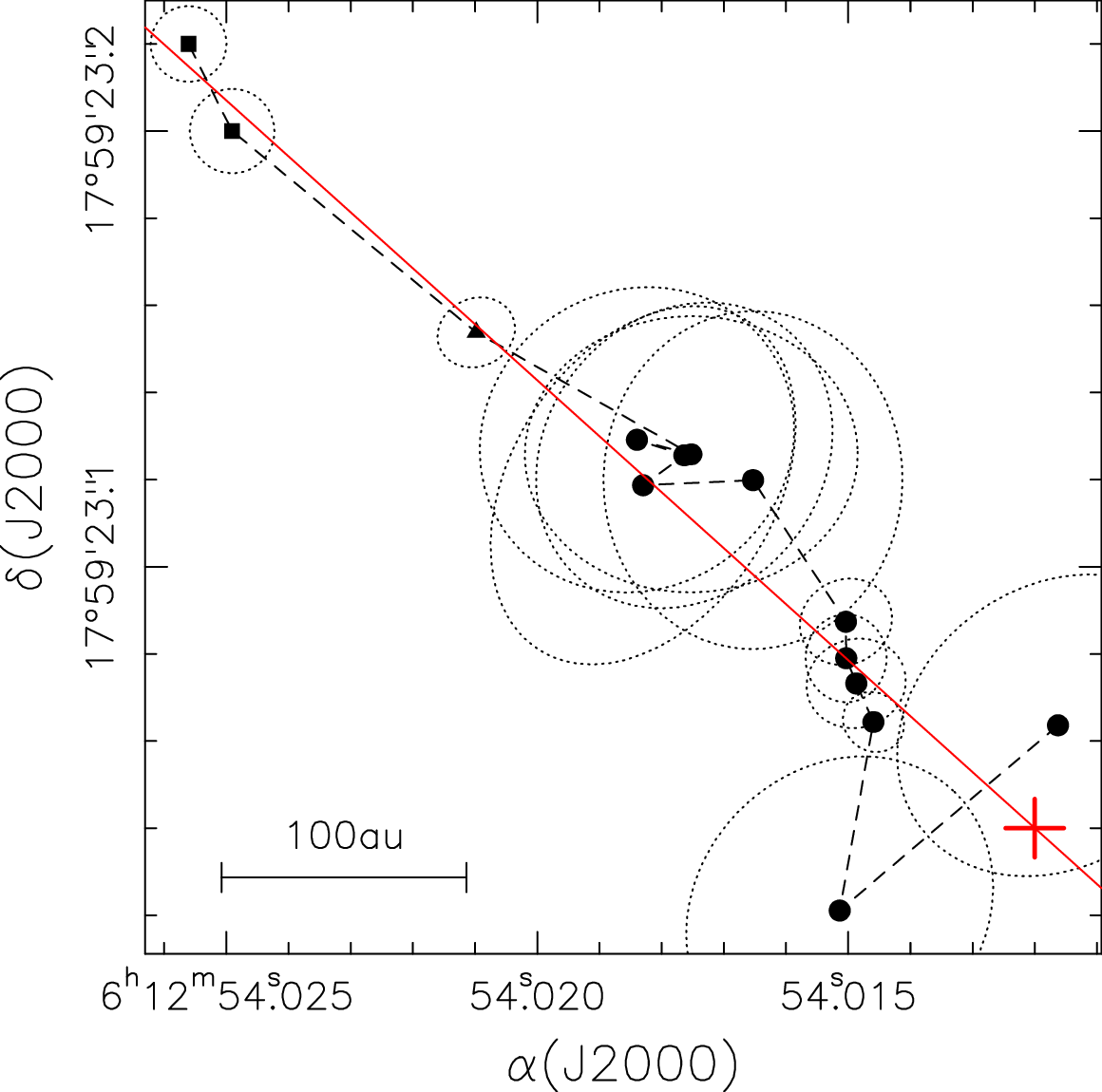}}
\caption{
Distribution of the peaks of the 3~mm and 1.3~cm continuum maps obtained,
respectively, with ALMA and with the A and B configurations of the VLA.
The circles indicate our VLA data, the triangle the data of OLHKP, and the
two squares our ALMA data. The dashed line connects the points in order of
time (the first point in the bottom-right corner corresponds to the observation on July 10,
2016). The third point from the top left is obtained from the OLHKP data. The
cross
marks the position of the (proto)star, and the solid
red line is the axis of the radio jet (PA=48\degr). The dotted ellipses are
the synthesised beams scaled down by a factor of 5.
}
\label{fpeaks}
\end{figure}

In Fig.~\ref{fsedfits} we present the continuum spectra obtained from the
flux densities in Table~\ref{tflux}.  Our VLA observations of \S\ span an
interval of time prior to the ALMA observations (see Table~\ref{tflux}) and
the radio emission of the variable source in \S\ is basically unresolved in
all of our VLA observations. However, we can study the spatial evolution of
the jet by determining the position of the peak at different times. For
this purpose we fitted a 2D Gaussian to the K-band maps with
sub-arcsecond resolution. We prefer the 1.3~cm data to the 7~mm
data, which would provide us with better resolution, because in the K band
the S/N is higher and in the Q band contamination by dust thermal emission
might be present. In Fig.~\ref{fpeaks} we plot the distribution of the
peak positions thus obtained. To give an idea of the uncertainty on these
positions, we also draw ellipses corresponding to one-fifth of the synthesised
beams. For our analysis we also included the NE peak of our ALMA data
and that of Obonyo et al.~(\cite{obonyo}; hereafter OLHKP). Despite the large
uncertainties, it is clear that the distance of the peak from the star is
increasing with time, as expected if the jet is expanding. This expansion
appears to slow down with time, because the mean velocity (projected on
the plane of the sky) estimated from the ratio between the separation
of the peaks at the last two epochs (ALMA data) and the corresponding
time interval is $\sim$40~au/820~days$\simeq$84~\kms, much less than
the mean velocity from the beginning of the radio burst, namely
$\sim$472~au/1881~days$\simeq$436~\kms, where $\sim$472~au is the
distance of the NE peak from the star at the last epoch. Using the same
approach, Fedriani et al.~(\cite{fedr23}) estimated an expansion speed of
$450\pm50$~\kms\ from their IR data. We further analyse the expansion
of the jet in Sect.~\ref{sexp}.

\begin{figure}
\centering
\resizebox{8.5cm}{!}{\includegraphics[angle=0]{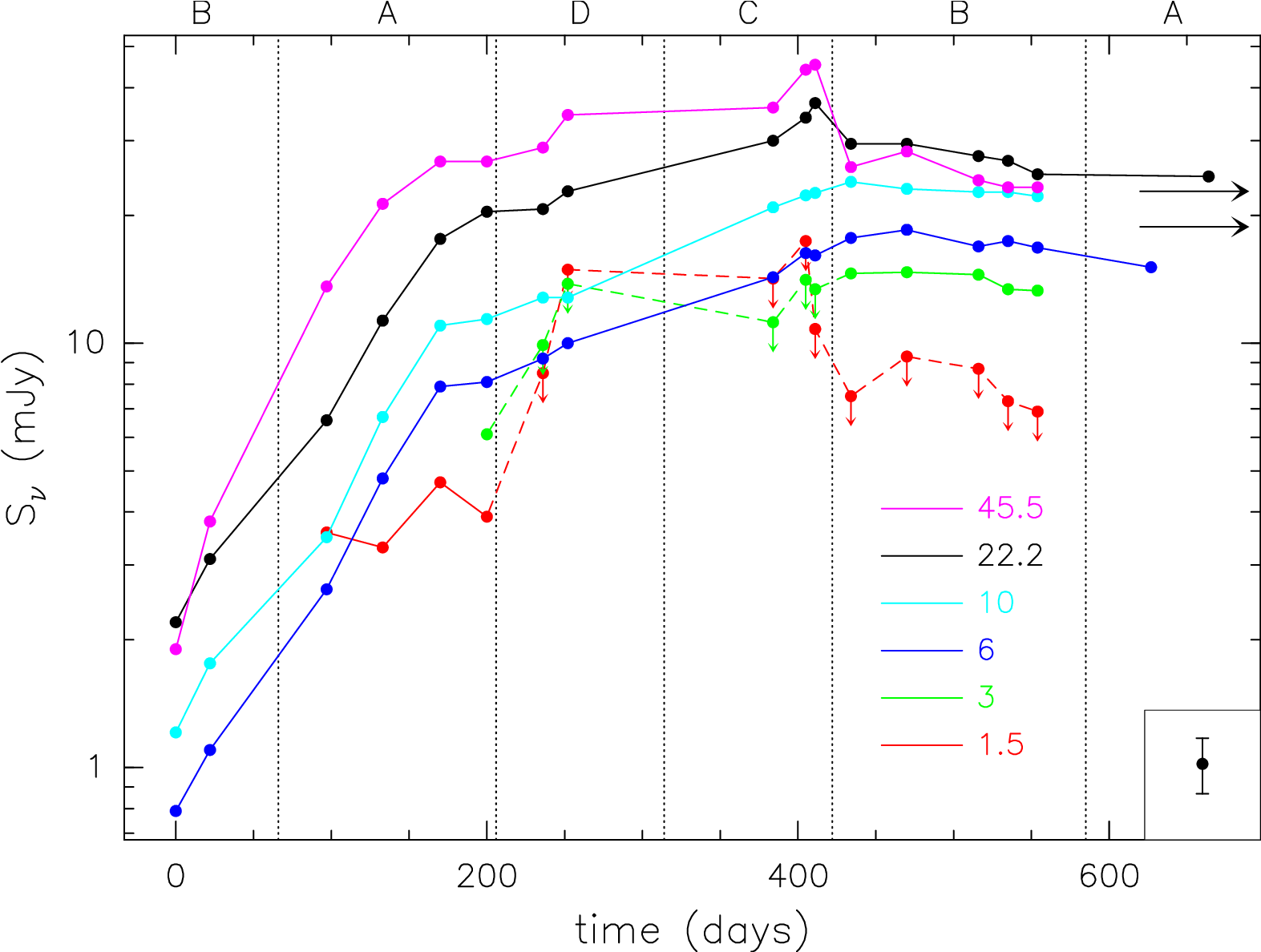}}
\caption{
Flux density of the variable, compact source in \S\ as a function of time
from the beginning of the radio burst (July 10, 2016) in all bands observed
with the VLA. The two rightmost points are the measurements at
6 and 22~GHz by OLHKP. Curves are colour-coded by frequency  (in gigahertz). Arrows denote points with upper limits (see
Sect.~\ref{svla}), which are connected by dashed lines. The two black
arrows pointing to the right indicate the flux densities at 92~GHz of
the NE lobe (see Table~\ref{tflux}) measured 1061 and 1881 days after
the radio burst. Typical error bars are shown in the bottom right of
the figure. The vertical dotted lines mark the beginning and end of an array
configuration, as labelled above the figure.
}
\label{fvar}
\end{figure}

The flux density of \S\ is changing with time in all bands,
as shown in Fig.~\ref{fvar}. One can identify two phases: the first
when the flux increases exponentially for $\sim$200~days; the second when
the flux remains basically constant, or slightly declines towards the
end of our monitoring. This behaviour is the same at all frequencies
$\ge$6~GHz, but is not so obvious at the two longest wavelengths, where
a precise estimate of the flux density of the compact variable source
is not possible for the reason explained in Sect.~\ref{svla}. Moreover,
in these bands our monitoring is more limited in time. It is hence quite
possible that the flux density below 6~GHz has a different behaviour than
that at higher frequencies. Therefore, the radio emission at 1.5 and 3~GHz,
and probably also part of the 6~GHz flux, might not be due to free-free
radiation but to another mechanism, such as synchrotron emission, which
has been detected towards a number of extended radio jets from YSOs
(e.g. Carrasco-Gonzalez et al.~\cite{cargon10}, Moscadelli et
al.~\cite{mosca13}, Brogan et al.~\cite{brog18}, Sanna et al.~\cite{sanna19}
and references therein). The
evident change of slope in some of the VLA spectra seems to support this
possibility. In this respect, the most representative is the spectrum
acquired on 2016/10/15 (see Fig.~\ref{fsedfits}), where the 1.5~GHz point
lies well above any plausible extrapolation of the other fluxes. For all
these reasons, in the following we focus our study on the emission
above 6~GHz, with the caveat that even the 6~GHz flux might be partly
contaminated by a non-thermal contribution, as suggested by OLHKP.
It is worth noting that the existence of synchrotron emission from \S\
is supported by the recent possible detection of high-energy gamma-ray
emission from this source (see Wilhelmi et al.~\cite{wilh23}).

\section{Analysis and discussion}

\subsection{Nature of the 3~mm continuum emission}
\label{smmem}

\begin{figure}
\centering
\resizebox{8.5cm}{!}{\includegraphics[angle=0]{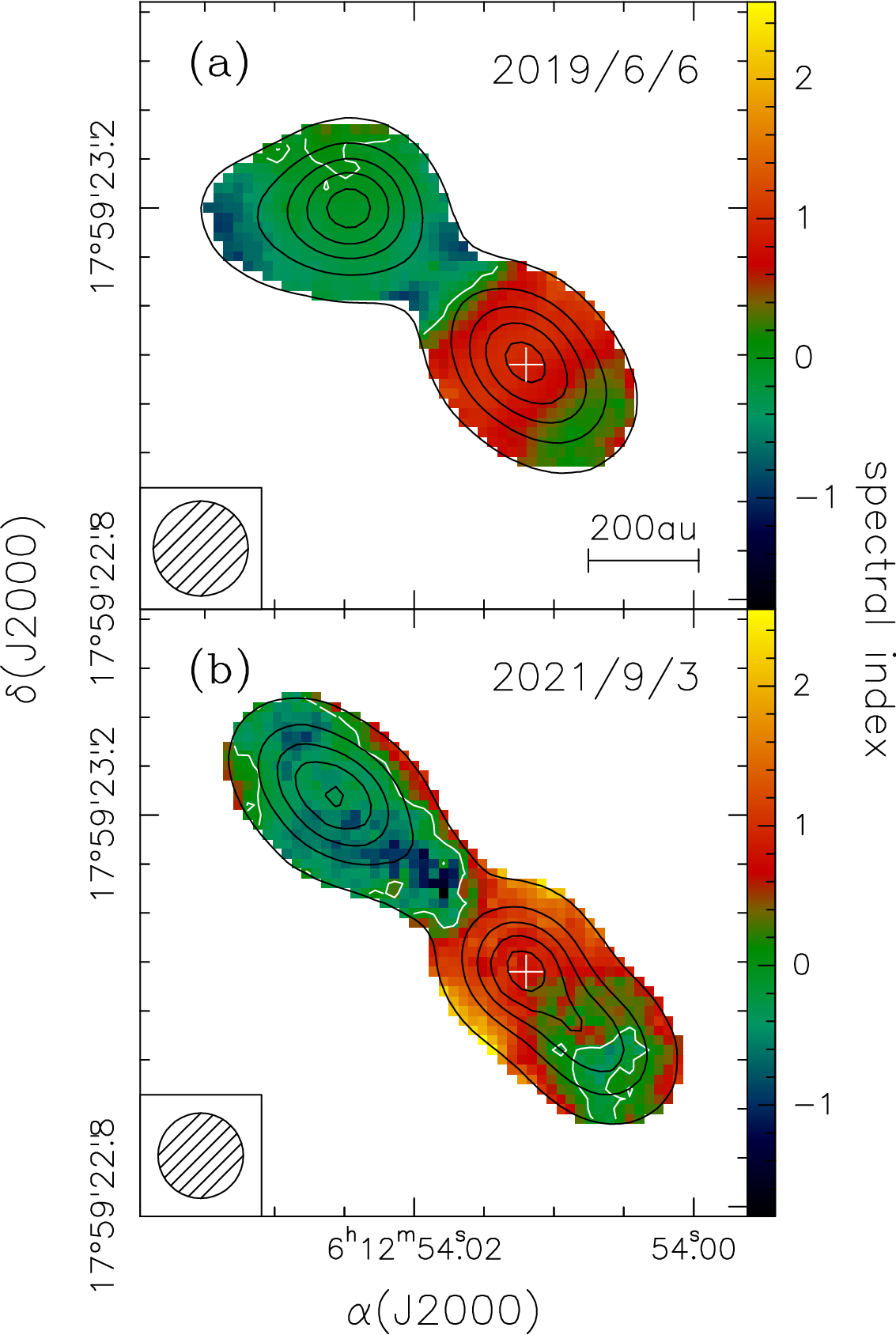}}
\caption{
Maps of the 3~mm continuum emission (contours) overlaid on the corresponding map of
the spectral index at 3~mm.
{\bf a.} Data obtained with ALMA on June 6, 2019.
Black contour levels range from 10\% to 90\% in steps of 10\% of the peak emission.
White contour levels correspond to a spectral index equal to 0.
The cross marks the position of the (proto)star.
The circle in the bottom left represents the synthesised beam.
{\bf b.} Same as the top panel, but for the map obtained on September 3, 2021.
}
\label{fspi}
\end{figure}

For the reasons presented in Sect.~\ref{srmm}, we have concluded
that the (proto)star should lie at the SW peak of the 3~mm continuum map.
If this is the case, one expects this peak to have a significant contribution
from thermal dust emission, whereas the NE peak, being part of a jet lobe,
should be dominated by free-free emission. To investigate these
assumptions, we computed the spectral index of the 3~mm continuum over the
maps in Fig.~\ref{falma}. For this purpose, we used the maps obtained from
the four correlator units centred at 85.2, 87.2, 97.2, and 99.2~GHz (see
Sect.~\ref{salma}), created with the same clean beam. For each pixel of
the maps, the spectral index was computed from a least-square fit to the
four fluxes in a $\log S_\nu$--$\log\nu$ plot. The fit was performed only
in those pixels where all of the four fluxes were above the 5$\sigma$ level.
The result is shown in Fig.~\ref{fspi}, where the formal error obtained
from the fit ranges from 0.03 towards the emission peaks to 0.3 towards
the borders.

At 3 mm it is reasonable to assume that dust emission is optically thin and
the flux density is $\propto\nu^\gamma$, with $\gamma=2$--4,
because the dust absorption coefficient is believed to vary as $\nu^\beta$
with $\beta=0$--2 (see e.g. D'Alessio et al.~\cite{dale01}, Sadavoy et
al.~\cite{sada13}), where $\beta=0$ corresponds to the case of large grains
(`pebbles') in disks (Testi et al.~\cite{testi}).
The same assumption also holds for the free-free emission: the spectra
in Fig.~\ref{fsedfits} flatten beyond 45 GHz, and hence the flux density
is $\propto\nu^{-0.1}$. Therefore, more positive spectral
indices can be associated with dust emission and, conversely, more negative
ones with free-free emission. At both epochs dust emission arises from the
region around the SW peak, as expected for a deeply embedded star, while the
NE lobe of the jet is characterised by free-free emission. Noticeably, some
free-free emission is also detected towards the most south-western tip.

When estimating the flux density at wavelengths of 3~mm or shorter,  it is thus
necessary to distinguish between the NE lobe and the rest of the source.
In Table~\ref{tf3mm} we give all these fluxes for the two epochs of the
ALMA observations. It is worth pointing out that the spectral index over
the SW region is mostly $\la$2, whereas the typical index expected for pure
dust emission, should lie approximately between 2 and 4. This suggests the
presence of non-negligible free-free emission around the SW peak as well. It
is possible to estimate what fraction of the total flux is due to free-free
as follows. The total flux can be written as $S_\nu=S_\nu^{\rm d}+S_\nu^{\rm
ff}$, where $S_\nu^{\rm ff}\propto\nu^{-0.1}$ is the optically thin free-free
flux and $S_\nu^{\rm d}\propto\nu^\gamma$, with $\gamma=2$--4, is the dust
flux. As previously explained, the spectral index between $\nu_1=85.2$~GHz
and $\nu_2=99.2$~GHz was estimated assuming $S_\nu\propto\nu^\alpha$,
and hence we have
\begin{eqnarray}
\frac{S_{\nu_2}}{S_{\nu_1}} & = & \left(\frac{\nu_2}{\nu_1}\right)^\alpha \\
\frac{S_{\nu_2}}{S_{\nu_1}} & = & \frac{S^{\rm d}_{\nu_2} + S^{\rm ff}_{\nu_2}}
                                   {S^{\rm d}_{\nu_1} + S^{\rm ff}_{\nu_1}} ~=~
                              \frac{S^{\rm d}_{\nu_1}\left(\frac{\nu_2}{\nu_1}\right)^\gamma + S^{\rm ff}_{\nu_1}\left(\frac{\nu_2}{\nu_1}\right)^{-0.1}}
                                   {S^{\rm d}_{\nu_1} + S^{\rm ff}_{\nu_1}}.
\end{eqnarray}
After some algebra, we obtain the ratio
\begin{equation}
 R_{\rm S} \equiv \frac{S^{\rm ff}_{\nu_1}}{S^{\rm d}_{\nu_1}}
             = \frac{\left(\frac{\nu_2}{\nu_1}\right)^\gamma - \left(\frac{\nu_2}{\nu_1}\right)^\alpha}
                    {\left(\frac{\nu_2}{\nu_1}\right)^\alpha - \left(\frac{\nu_2}{\nu_1}\right)^{-0.1}}
\end{equation}
and, from this, the fraction of the total flux due to free-free emission,
$R_{\rm S}/(1+R_{\rm S})$.  Using the values of $\alpha$ in Fig.~\ref{fspi},
we find that such a fraction on average ranges from 65\%, for $\gamma$=2,
to 85\%, for $\gamma$=4. This implies that 15--20~mJy out of 23~mJy emitted
by the SW component (see Table~\ref{tf3mm}), are contributed by free-free
emission, also consistent with the brightness
temperature of $\sim$500~K measured at 3~mm towards the SW peak, probably
too large to be due only to dust emission.
We note that part of this free-free emission might be due to ionisation
by the embedded star of $\sim$20~\Msun\ (Zinchenko et al.~\cite{zin15}).

\begin{table}
\centering
\caption[]{
Flux densities in millijanskys measured at 92.5~GHz with ALMA towards the NE and SW
components of \S.
}
\label{tf3mm}
\begin{tabular}{cccc}
\hline
\hline
date & NE & SW & total \\
\hline
 2019/6/6 & 22.8$^a$ & 23.2$^b$ & 46.0 \\
 2021/9/3 & 18.8$^a$ & 23.4$^b$ & 42.3 \\
\hline
\end{tabular}
\begin{flushleft}
$^a$~Free-free emission. \\
$^b$~Both dust and free-free emission.
\end{flushleft}
\end{table}
 
As already mentioned, some
free-free emission is also detected towards the tip of the SW
region and becomes more prominent and more extended with time, as one can
see by comparing Fig.~\ref{falma}a to~Fig.~\ref{falma}b. This hints
at the existence of another jet lobe emerging from the dusty core and
expanding towards the SW.

\begin{figure}
\centering
\resizebox{8.5cm}{!}{\includegraphics[angle=0]{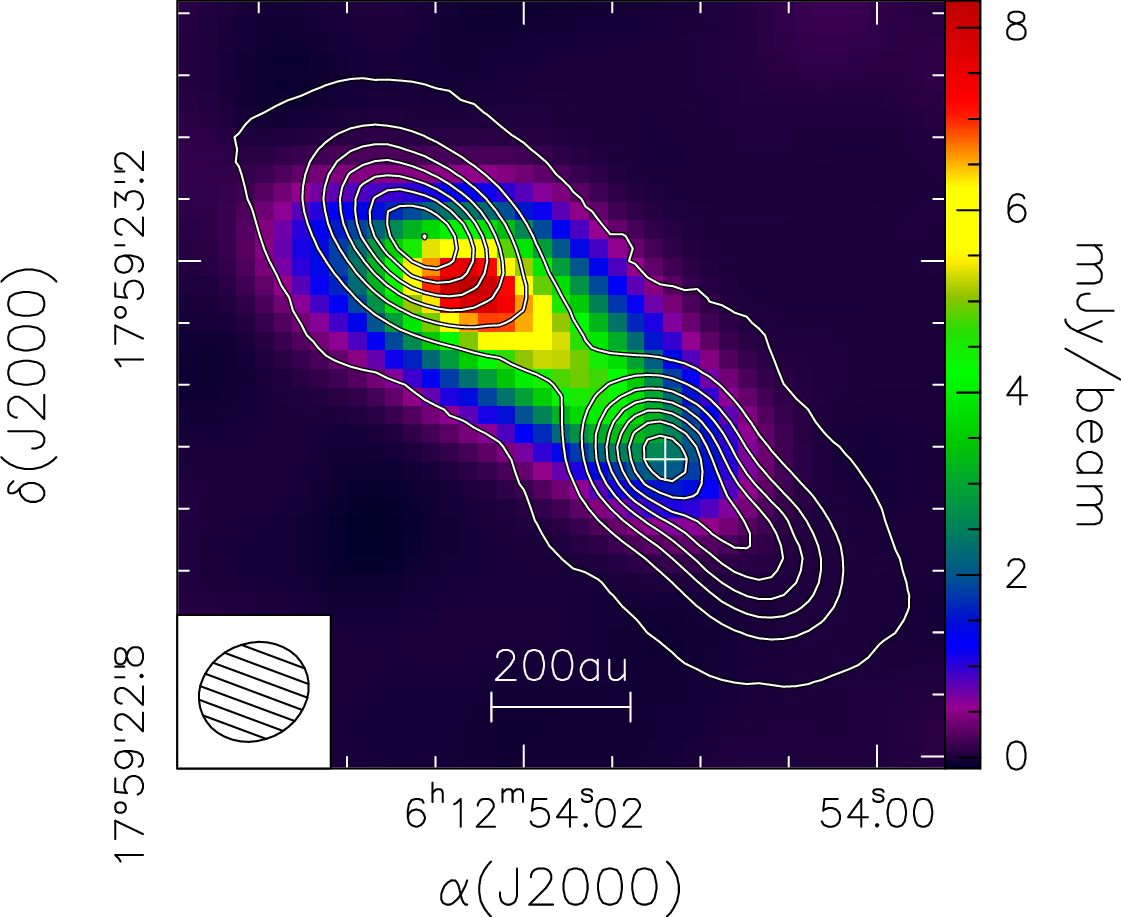}}
\caption{
Map of the 3~mm continuum emission obtained with ALMA on
September 3, 2021,
overlaid on the 1.3~cm image obtained by OLHKP
on May 5, 2018. Contour levels are the same as in
Fig.~\ref{falma}b. The cross marks the position of the (proto)star.
The
circle in the bottom left represents the synthesised beam at 1.3~cm.
}
\label{flobes}
\end{figure}

To shed light on the nature of this putative lobe and set a constraint on its
age, in Fig.~\ref{flobes} we compare the jet structure observed by OLHKP at 1.3~cm on May 5, 2018, with
our ALMA image at 3~mm.  It seems that during the period of our monitoring,
up to the observation of OLHKP (664 days after the onset of the radio
outburst), no significant
free-free
emission was seen to the SW of the (proto)star
at any of the wavelengths observed with the VLA.
We
thus conclude that in all likelihood the SW jet lobe appeared only recently,
between May 2018 and June 2019. Therefore, the radio flux variations
monitored by us until January 2018 are to be attributed only to the NE lobe.

\subsection{Origin of the SW lobe}

One may wonder if the emerging SW lobe corresponds to a new radio burst or
is somehow related to the accretion outburst observed by Caratti o Garatti
et al.~(\cite{cagana}). Only follow-up observations of the jet structure
will allow us to establish if the new lobe is as prominent and long-lasting
as the NE one. However, we point out that the IR monitoring performed by
Uchiyama et al.~(\cite{uchi}) and Fedriani et
al.~(\cite{fedr23}) between November 2015 and February 2022
as well as the methanol maser observations\footnote{Available at
http://vlbi.sci.ibaraki.ac.jp/iMet/data/192.6-00} of the Maser Monitoring
Organization (M2O)\footnote{The M2O is a global cooperative of maser
monitoring programmes; see https://MaserMonitoring.org} have not revealed
any other burst after that of Caratti o Garatti et al.~(\cite{cagana})
and before the ejection of the SW lobe. We conclude that both jet lobes
could arise from the same accretion event, although with a time lag between
them of 22--35~months.

\begin{figure} 
\centering
\resizebox{8.5cm}{!}{\includegraphics[angle=0]{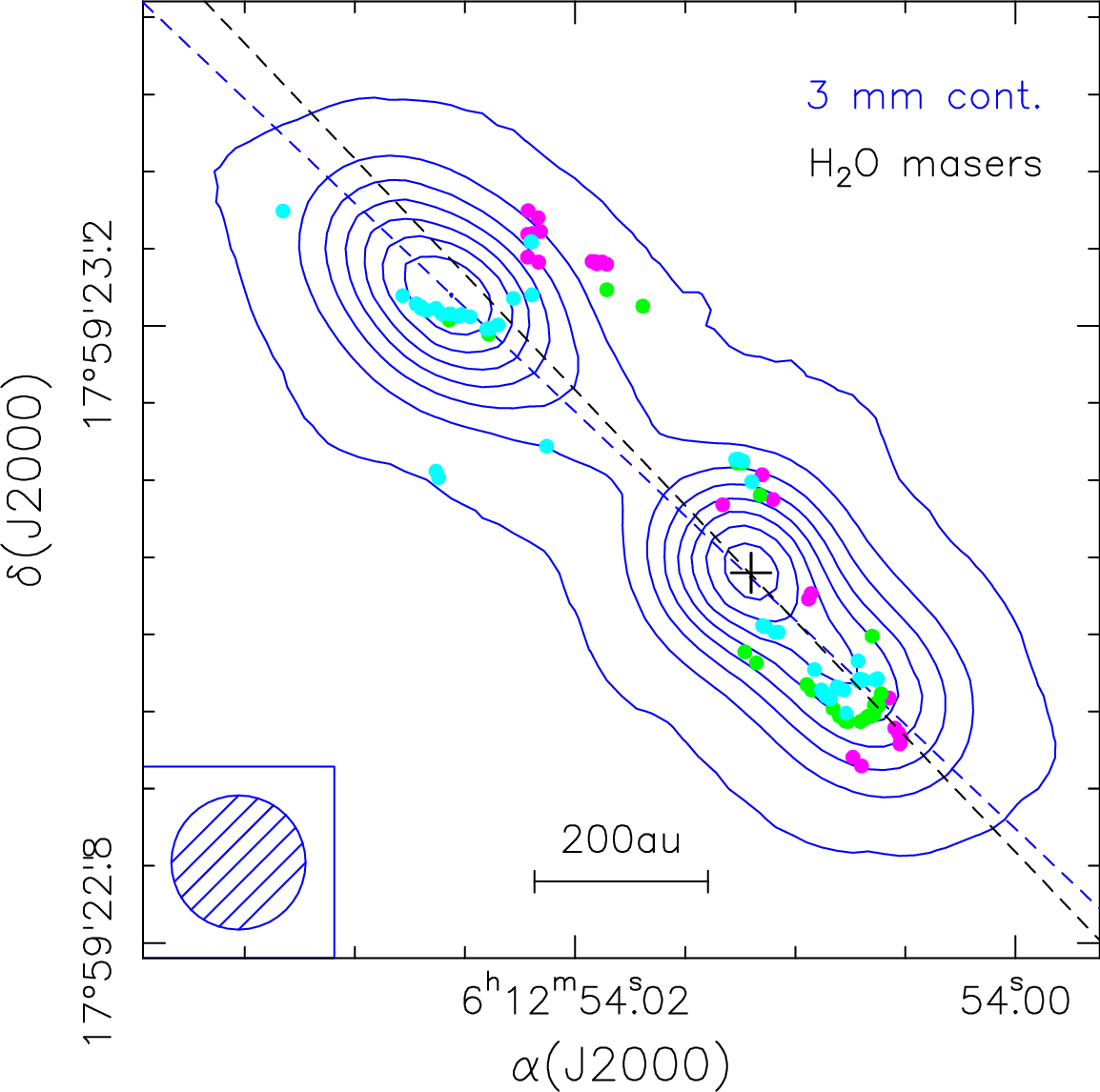}}
\caption{
Water masers spots (solid circles) observed by Goddi et al.~(\cite{goddi07}; cyan), Burns et al.~(\cite{burns16}; green), and
Hirota et al.~(\cite{hiro21}; magenta) overlaid on the 3~mm continuum map (blue contours) of Fig.~\ref{falma}b. The
dashed lines indicate the jet axes in the two tracers. The cross marks
the position of the (proto)star. The circle in the bottom left represents
the synthesised beam at 3~mm.
}
\label{fwat}
\end{figure}

This hypothesis is also supported by a noticeable feature of the
jet system in \S, namely that the extension of the NE lobe is about
twice as much as that of the SW lobe. This is true not only for the
large-scale and small-scale lobes in Fig.~\ref{fjets}, but also for the
\WAT\ masers distribution, as shown in Fig.~\ref{fwat}. The existence
of the same asymmetry on different
scales and tracers cannot be a coincidence and is suggestive of a mechanism
that causes a delay of the ejection of the SW lobe with respect to the NE
lobe. In our opinion, two explanations are possible: either the accretion
(and hence ejection) event is intrinsically stronger on the NE side of the
disk, or the expansion towards the SW is hindered by the presence of denser
material. We favour the latter hypothesis since the near-IR emission is
much fainter from the SW lobe than from the NE lobe (see Caratti o Garatti
et al.~\cite{cagana} and Fedriani et al.~\cite{fedr23}). This finding is
surprising, because the SW lobe corresponds to the blue-shifted emission
of the jet outflow (see Wang et al.~\cite{wang}), which means that the jet
is pointing towards the observer on that side and the extinction should
be lower than on the NE side. The weakness of the IR emission to the SW is
thus indicative of an asymmetry in the density distribution along the jet
axis, a fact that could naturally also explain the delay of the expansion of
the SW lobe with respect to the NE lobe.

\subsection{Evolution of the radio emission}

In this section we present a model fit to the observed spectra, following
the approach adopted in Paper I, with some modifications. As done in
Paper~I, we adopted the `standard spherical' model from Reynolds~(\cite[see his Table~1]{reyn}) to describe the radio continuum emission from the jet. In
practice, this means that we assume a jet where at a given time the opening
angle, electron temperature, ionisation degree, and expansion speed do
not depend on the distance from the star, $r$.

\subsubsection{Expansion law of the jet}
\label{sexp}

In Paper~I the jet was assumed to undergo expansion at constant velocity
so that the maximum radius could be described by the simple expression
$\rmax(t)=\rmax(0)+\vo t$, with $\vo$=900~\kms. While this assumption
could hold for the first few months after the onset of the radio outburst,
it is inconsistent with the most recent data. In fact, in Sect.~\ref{srcm}
we show that the jet expansion is slowing down with time. We thus
need to adopt a more realistic law for $\rmax(t)$.

For this purpose, we assume that the jet is expanding in a medium with
density $\propto r^{-2}$, with $r$ the distance from the star. It is possible
to demonstrate (see Appendix~\ref{aexp}) that applying momentum
conservation one obtains
\begin{equation}
 \ym(t) = \ymo + 2T\vo\cos\psi \left(\sqrt{1+\frac{t}{T}}-1\right)  \label{eym}
,\end{equation}
where we have multiplied both terms of Eq.~(\ref{earm}) by $\cos\psi$, with
$\psi$ the angle between the jet axis and the plane of the sky. For consistency
with Reynolds~(\cite{reyn}), we have indicated with $y$ the projection
of $r$ on the plane of the sky. Here, 
$T$ is a suitable timescale, and $\vo$ and $\ymo$ are, respectively,
the expansion velocity and the value of $\ym$ at the onset of the radio
outburst (i.e. on July 10, 2016). As in Paper~I, we chose $\vo$=900~\kms,
while the two parameters $T$ and $\ymo$ can be determined from the values
of $\ym$ estimated from the two ALMA maps at $t$=1061~days and $t$=1881~days.

The problem is that $\ym$ cannot be trivially obtained from the position
of the NE peak, which corresponds to the maximum brightness temperature.
This temperature is attained either at the inner radius, if the whole jet is
optically thin, or at the border between the optically thin and the optically
thick parts of the jet (denoted by $y_1$ in Reynolds' notation). Beyond this
point, the emission is optically thin and the brightness temperature scales
with the opacity $\tau \propto r^{-3}$, as from Reynolds' Eq.~(4). For this
reason, the jet can extend much beyond the position of the observed peak.

\begin{figure}
\centering
\resizebox{8.5cm}{!}{\includegraphics[angle=0]{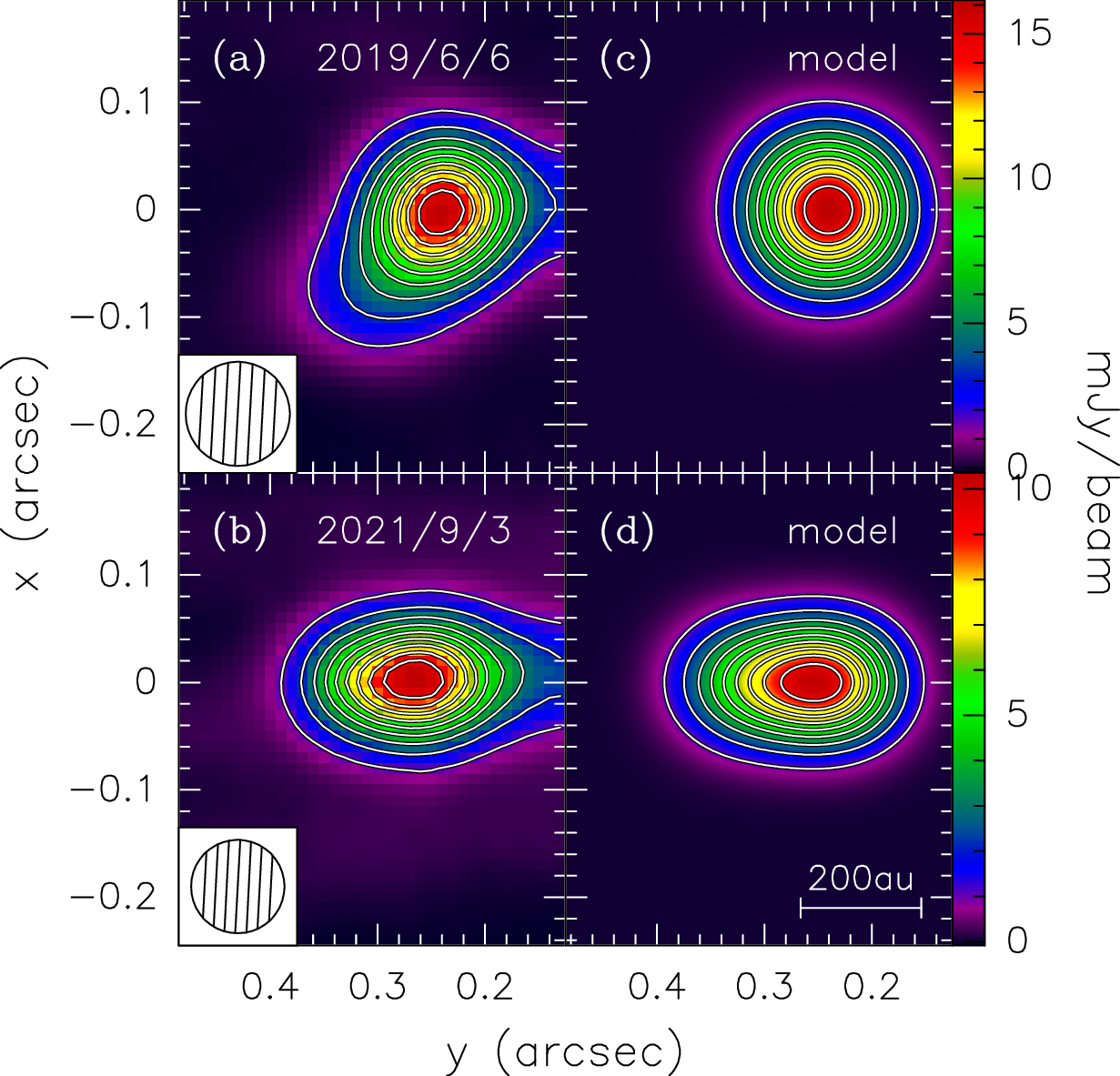}}
\caption{
Comparison between the observed maps of the NE lobe of the radio jet (left
panels) with the best-fit maps obtained with the model (right panels). The top
panels correspond to the first ALMA epoch, bottom panels to the second
one. The horizontal axis is parallel to the jet axis. Contour levels range
from 10\% to 90\% in steps of 10\% of the peak of each image. The circle
in the bottom left represents the synthesised beam.
}
\label{fmods}
\end{figure}

In order to obtain a reliable estimate of $\ym$, we have fitted the NE lobe
in the two ALMA maps of Fig.~\ref{falma} assuming that the 3~mm emission is
optically thin all over the jet surface. Under this hypothesis the brightness
temperature can be expressed as
\begin{equation}
T_{\rm B} = T_0 \tau = T_0 \left(\frac{y}{\yo}\right)^{-3}
,\end{equation}
where $y=r\,\cos\psi$ is the projection of $r$ on the plane of the sky and
$\yo=\ro\,\cos\psi$ with $\ro$ the inner radius of the jet. For a given
set of $\yo$, $\ym$, and $\tho$ (the opening angle of the jet) a map was
generated, convolved with the instrumental beam, and the model brightness
temperature was computed for each pixel of the observed map. The best fit
was obtained by minimising the expression
\begin{equation}
\chi^2=\sum_i \left(T_{\rm B}^i({\rm model})-T_{\rm B}^i({\rm data})\right)^2
,\end{equation}
with $i$ a generic pixel,
after varying the three parameters over suitable ranges. The best-fit models
are compared to the observed maps in Fig.~\ref{fmods} and correspond to
$\tho$=9\degr, $\yo$=0\farcs206=367~au, and $\ym$=0\farcs29=516~au, for the first map, and
$\tho$=2\fdg4, $\yo$=0\farcs207=368~au, and $\ym$=0\farcs356=634~au, for the second map.

\begin{figure}
\centering
\resizebox{8.5cm}{!}{\includegraphics[angle=0]{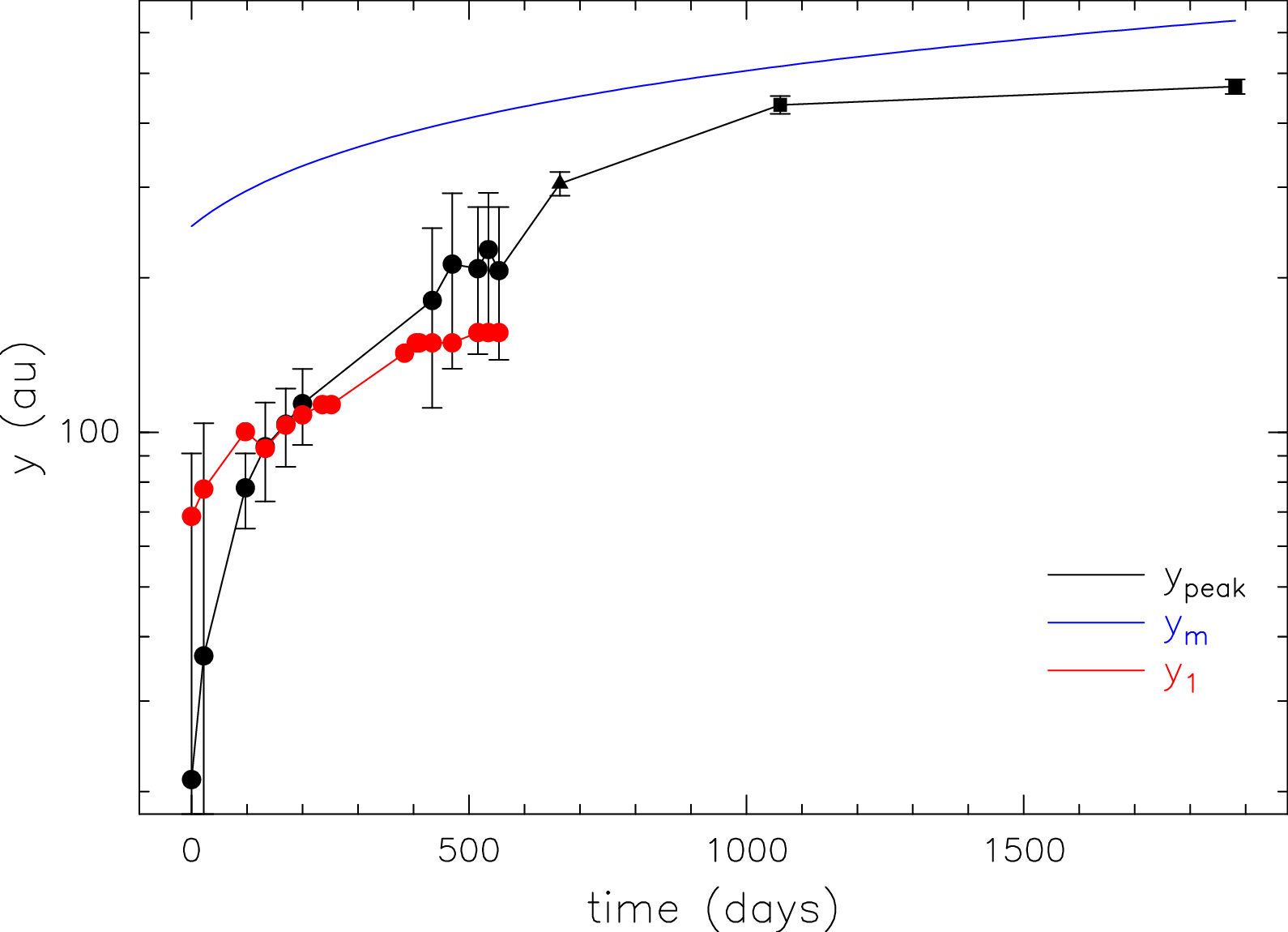}}
\caption{
Comparison of the projection on the plane of the sky of the outer radius
of the jet, $\ym$ (blue), with the position of the brightness temperature peak,
$y_{\rm peak}$ (black), and the border between the optically thin and optically
thick parts of the jet, $y_1$ (red). All quantities are plotted as a function
of time from the onset of the radio outburst (July 10, 2016). The black
circles indicate our VLA data, the triangle the OLHKP data, and the
two squares our ALMA data. We remark that $y_{\rm peak}$ is measured from
Fig.~\ref{fpeaks} as the separation between the star and the projection
of the peak along the jet axis.
}
\label{frmax}
\end{figure}

From Eq.~(\ref{eym}) written for $t$=1061~days and $t$=1881~days, one
obtains a system of two equations in the two unknowns $T$ and $\ymo$, the
solutions to which are $T$=124~days and $\ymo$=251~au. Hence, one has
\begin{equation}
 \ym({\rm au}) = 251 + 127 \left(\sqrt{1+\frac{t({\rm days})}{124}}-1\right). \label{eymnum}
\end{equation}
The value of $\rmax$
can be computed from Eq.~(\ref{eym}) assuming $\psi$=10\degr\
(see Paper~I, where the complementary angle \mbox{$i=90\degr-\psi=80\degr$}
was used). Figure~\ref{frmax} compares our solution, $\ym$, as a function of time (blue curve) 
with (i) the positions of the maximum brightness temperature (the same as in
Fig.~\ref{fpeaks}) and (ii) the value of $y_1$ obtained from the model fits
described later in Sect.~\ref{smod}. The latter corresponds to the border
between the optically thin and optically thick parts of the jet. Clearly,
the brightness appears to peak much closer to $y_1$ than to $\ym$, as previously
mentioned.

\subsubsection{Modelling the jet variability}
\label{smod}

We wish to reproduce the spectral variation shown in Fig.~\ref{fsedfits}
and thus derive the values of the jet parameters as a function of time. For
this purpose, we introduce some modifications with respect to the original model
by Reynolds~(\cite{reyn}), which was used in Paper~I. Reynolds' equations
were derived under the approximation of small $\tho$, the opening
angle of the jet\footnote{We stress that our definition of $\tho$ corresponds
to one-half of the $\tho$ defined by Reynolds~(\cite{reyn}).}. However,
in Paper~I we find that $\tho$$\simeq$20\degr--50\degr\
is needed to fit the spectra. In order to overcome this limitation we re-wrote Reynolds' equations under suitable assumptions, as detailed
in Appendix~\ref{amodel}, so that they are now valid for any $\tho<90\degr$.

All observed spectra have been fitted with Eq.~(\ref{eaflux}) using
only the measurements with $\nu$$\ge$6~GHz, for the reasons discussed in
Sect.~\ref{srcm}. We stress that, unlike Paper~I, here we assume the jet
to be mono-polar, because there is no hint of the existence of a SW lobe
during the whole period of our VLA monitoring (see Sect.~\ref{smmem}). The
input parameters of the model are the angle between the jet axis and the
plane of the sky, $\psi$, the ionised gas temperature, $T_0$, the inner
radius, $\ro$, the projection of the outer radius on the plane of the sky,
$\ym$, the opening angle, $\tho$, and the parameter $\MM=x_0\,\dot{M}/\vj_0$,
where $x_0$ is the fraction of ionised gas, $\vj_0$ the expansion velocity,
and $\dot{M}$ the total mass loss rate (neutral plus ionised). The quantities
$T_0$, $\vj_0$, and $x_0$ are assumed to be constant along the jet, while
$T_0$ is also constant in time.

\begin{figure}
\centering
\resizebox{8.5cm}{!}{\includegraphics[angle=0]{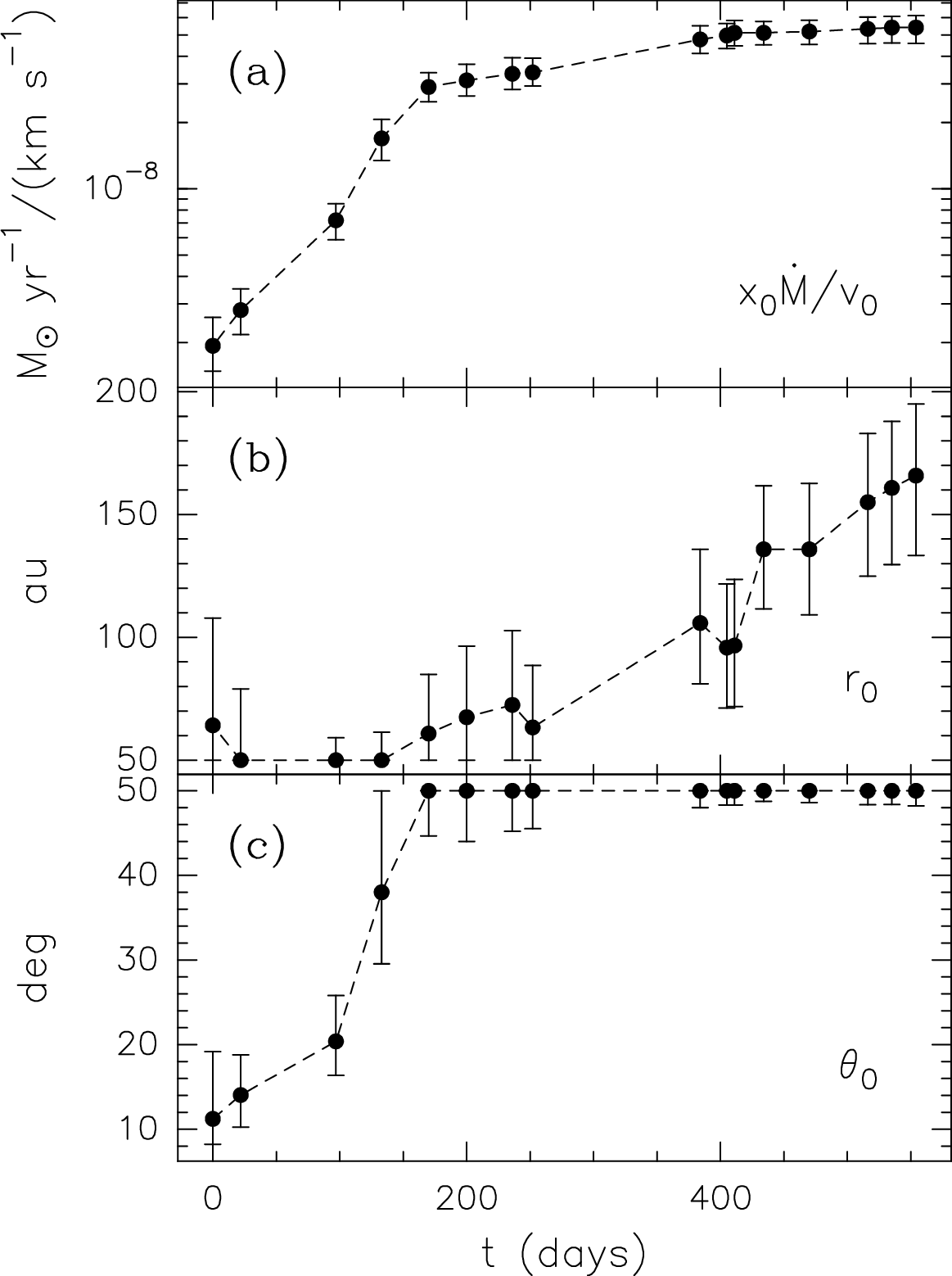}}
\caption{
Plot of the parameters of the best fits to the continuum spectra,
as a function of time from the onset of the radio outburst (July~10, 2016).
The plotted parameter is indicated in the bottom right of each panel.
}
\label{fpars}
\end{figure}

\begin{figure}
\centering
\resizebox{8.5cm}{!}{\includegraphics[angle=0]{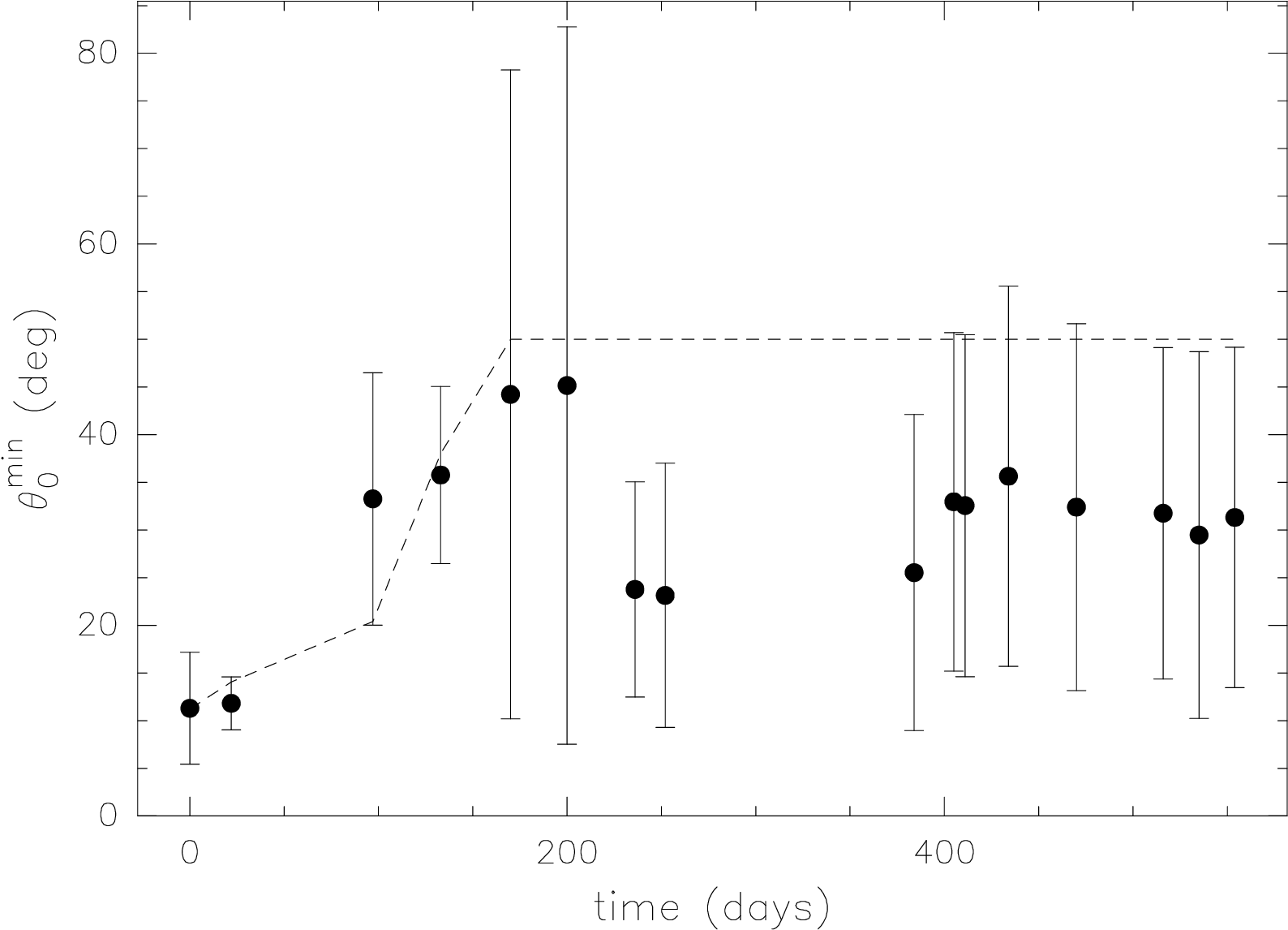}}
\caption{
Lower limit on the opening angle estimated from Eq.~(\ref{etho}) for all the
epochs of our monitoring. The bars indicate the standard deviation of
the values obtained in the different bands. The dashed line connects the values
of $\tho$ obtained from the best fits to the spectra (see Table~\ref{tpars}).
}
\label{ftho}
\end{figure}

To simplify the fitting procedure as much as possible,
we fixed $T_0=10^4$~K and $\psi=10\degr$ (see Paper I), and
computed $\ym$ from Eq.~(\ref{eymnum}). Unlike in Paper~I, we
decided to leave $\ro$ free, because a priori the inner radius
could change while the jet is expanding. So, we are left with
three free parameters: $\ro$, $\tho$, and $\MM$.
The best fit to each spectrum has been obtained by minimising
the $\chi^2$ given in Eq.~(10) of Paper~I, after varying the
parameters over the ranges $\tho$=3\degr--50\degr, $\ro$=50--300~au, and
$\MM$=$10^{-10}$--$10^{-7}$~\Msun~yr$^{-1}$/(\kms).
The best-fit spectra are represented by the solid curves in
Fig.~\ref{fsedfits} and the best-fit parameters are given in
Table~\ref{tpars}. The errors on the parameters have been computed
using the criterion of Lampton et al.~(\cite{lamp}), as done in Paper~I.

While most of the fits look to be in agreement with the data within the
uncertainties, the 6~GHz fluxes appear to be underestimated
by the model at the last five to six epochs. In our opinion, such a discrepancy could indicate contamination from non-thermal emission at this frequency, as already
suggested by OLHKP. Indeed, as discussed in Sect.~\ref{srcm}, non-thermal
emission is very prominent at longer wavelengths in the same spectra (red
points in Fig.~\ref{fsedfits}) and it is hence not surprising that this
type of emission can contribute significantly to the flux up to 6~GHz.

In Fig.~\ref{fpars} we plot the best-fit parameters as a function of
time. One sees that the opening angle rapidly increases up to the maximum
value allowed by us. It may seem that an opening angle of 50\degr\ is
too large for a jet, but it is similar to that predicted by theory for
a jet powered by a $\sim$20~\Msun\ YSO (Zinchenko et al.~\cite{zin15}),
namely $\sim$52\degr, obtained by interpolating the values in Table~2 of
Staff et al.~(\cite{staff19}).
Moreover, we can obtain a direct estimate of $\tho$ from the observed peak
brightness temperature in the synthesised beam, $\Tsb$, assuming optically
thick emission. Approximating the jet as a Gaussian source, one has
\begin{equation}
\Tsb=\Tb\frac{\Theta_{\rm S}^2}{\Theta_{\rm S}^2+\Theta_{\rm B}^2}
,\end{equation}
where $\Tb$ is the intrinsic brightness temperature of the source and
$\Theta_{\rm S}$ and $\Theta_B$ are, respectively, the full widths at half
power of the source and synthesised beam. Consequently,
\begin{equation}
\Theta_{\rm S} = \Theta_{\rm B} \sqrt{\frac{\Tsb}{\Tb-\Tsb}}. \label{eths}
\end{equation}
This expression can be used to calculate a lower limit on the source diameter,
$\Theta_{\rm S}^{\rm min}$, assuming that the free-free
emission is optically thick, namely for $\Tb$=$\To$=$10^4$~K. Correspondingly,
a lower limit on the opening angle is obtained from
\begin{equation}
\tho^{\rm min} = \arcsin\left(\frac{\Theta_{\rm S}^{\rm min}}{2\Delta}\right) = \arcsin\left(\frac{\Theta_{\rm B}}{2\Delta}\sqrt{\frac{\Tsb}{\To-\Tsb}}\right)   \label{etho}
,\end{equation}
where $\Delta$ is the separation between the peak of $\Tsb$ and the position
of the (proto)star. We estimated $\tho^{\rm min}$ for all of our maps, and for
each epoch we computed the mean $\tho^{\rm min}$ obtained from the four bands. This is plotted in
Fig.~\ref{ftho} as a
function of time. For the sake of comparison we also plot $\tho$ obtained
from our model fits (dashed line). We conclude that, despite the crude approximations
adopted to derive Eq.~(\ref{etho}), values of $\tho$ of a few times 10\degr\
seem plausible and consistent with our model fit results. It is worth
noting that $\tho$ computed from the map of OLHKP and from our ALMA maps
(Fig.~\ref{falma}) using Eq.~(\ref{etho}) turns out to be much less,
namely 7\degr, 1\fdg7, and 1\fdg3, respectively 664, 1061, and 1881 days
after the radio outburst. We speculate that the jet could be
re-collimating on a timescale of a couple of years.

\begin{table}
\caption[]{
Parameters of the best fits to the spectra in Fig.~\ref{fsedfits}.
}
\label{tpars}
\renewcommand{\arraystretch}{1.3}
\begin{tabular}{cccccc}
\hline
\hline
date & time & $\ym$$^a$ & $\theta_0$ & $\ro$ & $\MM\times10^8$ \\
{\tiny (yy/mm/dd)} & {\tiny (days)} & (au) & (deg) & (au) &
{\begingroup\makeatletter\def\f@size{8}\check@mathfonts
$(M_\odot~{\rm yr}^{-1}/({\rm km~s}^{-1}))$\endgroup} \\
\hline
16/07/10 &   0 & 251 & $11_{-3}^{+8}$ &  $64_{-14}^{+44}$ & $0.19_{-0.04}^{+0.07}$ \\
16/08/01 &  22 & 262 & $14_{-4}^{+5}$ &  $50_{-0}^{+29}$ & $0.28_{-0.06}^{+0.07}$ \\
16/10/15 &  97 & 294 & $20_{-4}^{+5}$ &  $50_{-0}^{+9}$ & $0.72_{-0.13}^{+0.14}$ \\
16/11/20 & 133 & 307 & $38_{-8}^{+12}$ &  $50_{-0}^{+11}$ & $1.69_{-0.35}^{+0.37}$ \\
16/12/27 & 170 & 320 & $50_{-5}^{+0}$ &  $61_{-11}^{+24}$ & $2.90_{-0.41}^{+0.47}$ \\
17/01/26 & 200 & 329 & $50_{-6}^{+0}$ &  $68_{-18}^{+29}$ & $3.11_{-0.47}^{+0.57}$ \\
17/03/03 & 236 & 340 & $50_{-5}^{+0}$ &  $73_{-23}^{+30}$ & $3.33_{-0.50}^{+0.61}$ \\
17/03/19 & 252 & 345 & $50_{-4}^{+0}$ &  $63_{-13}^{+25}$ & $3.38_{-0.45}^{+0.54}$ \\
17/07/29 & 384 & 381 & $50_{-2}^{+0}$ & $106_{-25}^{+30}$ & $4.78_{-0.65}^{+0.73}$ \\
17/08/19 & 405 & 386 & $50_{-2}^{+0}$ &  $96_{-25}^{+26}$ & $4.98_{-0.63}^{+0.66}$ \\
17/08/25 & 411 & 388 & $50_{-2}^{+0}$ &  $97_{-25}^{+27}$ & $5.12_{-0.66}^{+0.71}$ \\
17/09/17 & 434 & 393 & $50_{-1}^{+0}$ & $136_{-24}^{+26}$ & $5.12_{-0.60}^{+0.65}$ \\
17/10/23 & 470 & 402 & $50_{-1}^{+0}$ & $136_{-27}^{+27}$ & $5.19_{-0.66}^{+0.65}$ \\
17/12/08 & 516 & 413 & $50_{-2}^{+0}$ & $155_{-30}^{+28}$ & $5.33_{-0.76}^{+0.71}$ \\
17/12/27 & 535 & 417 & $50_{-2}^{+0}$ & $161_{-31}^{+27}$ & $5.41_{-0.81}^{+0.67}$ \\
18/01/15 & 554 & 421 & $50_{-2}^{+0}$ & $166_{-32}^{+29}$ & $5.41_{-0.83}^{+0.72}$ \\
\hline
\end{tabular}
\begin{flushleft}
$^a$~obtained from Eq.~(\ref{eymnum})
\end{flushleft}
\end{table}

\begin{figure}
\centering
\resizebox{8.5cm}{!}{\includegraphics[angle=0]{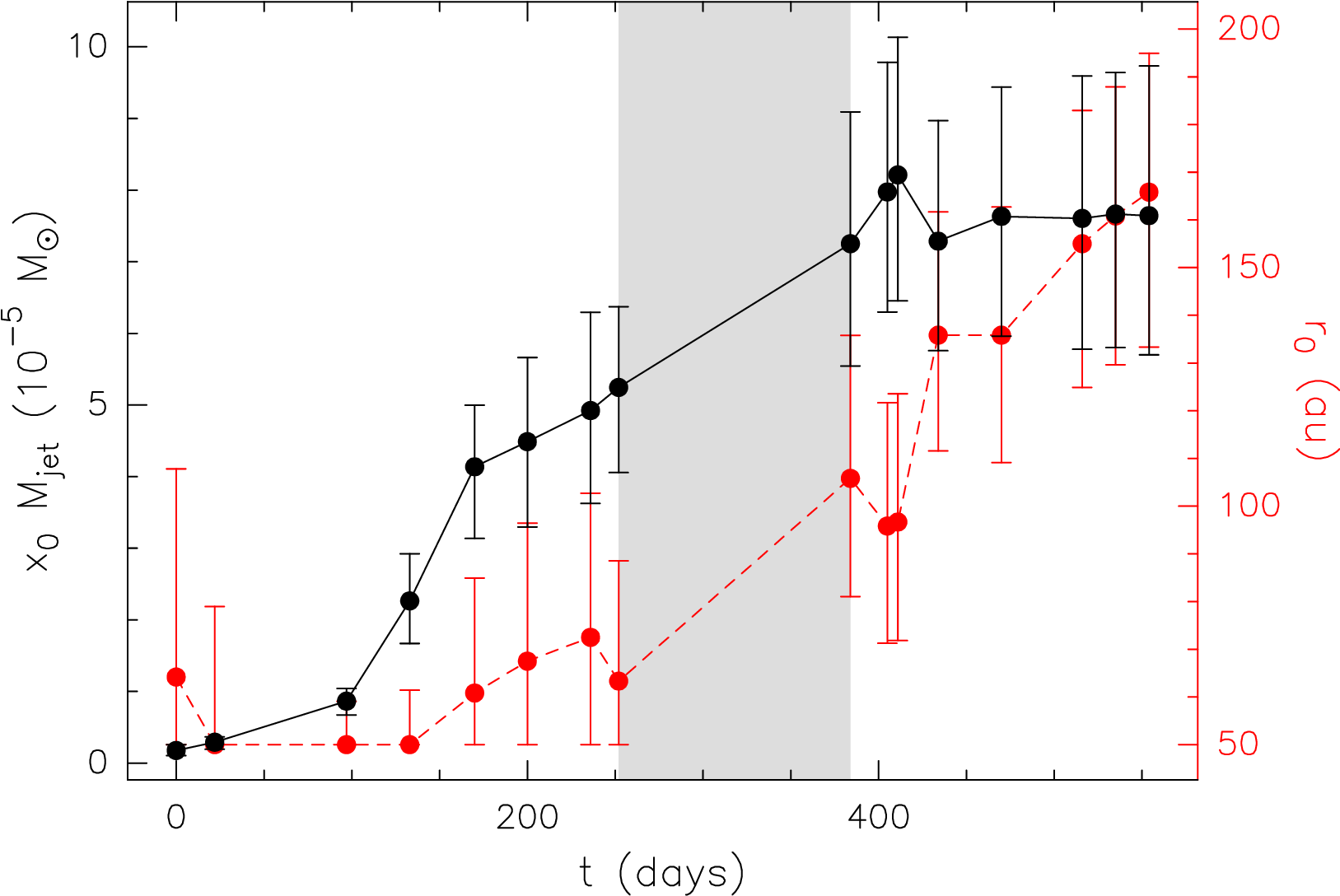}}
\caption{
Plot of the jet mass (solid black curve) and inner radius (dashed red curve)
as a function of time from the onset of the radio outburst (July~10, 2016).
The grey area separates the period when the jet is still
powered by the outburst from that when the feeding mechanism is quenched.
}
\label{fmjet}
\end{figure}

Another interesting result from Fig.~\ref{fpars} is the behaviour of
$\ro$, which remains basically constant for $\sim$250~days after the radio
outburst, thereafter showing a systematic increase until the end of our monitoring
-- and even beyond that, as we estimate values of $\sim$370~au at the time
of our first ALMA observations (see Sect.~\ref{sexp}). A straightforward
interpretation is that the inner radius expands when the mechanism feeding
the jet is switched off.  Noticeably, $\MM$ appears to reach a maximum just
when $\ro$ starts increasing, supporting the idea that no more material
is added to the jet. This hypothesis is confirmed by Fig.~\ref{fmjet},
which shows how the ionised jet mass, computed from Eq.~(\ref{eamjet}),
varies with time. In the same figure we also plot $\ro$ for the sake
of comparison. It is quite clear that both quantities have a bi-modal
temporal behaviour, before and after the time interval marked by the grey
area. From the onset of the radio outburst up to $\sim$250~days, $\ro$
remains approximately constant, whereas the jet mass increases. After
that, the reverse occurs: as soon as the inner radius starts increasing,
the jet stops growing in mass, with the grey area marking the transition
between these two phases. This is consistent with mass conservation in an
expanding jet that is not fed anymore by the outburst.

It is also worth noting that the ratio, $R_{\rm e}$, between the
mass ejected, $M_{\rm jet}$, and that accreted during the outburst,
$M_{\rm acc}$, can be computed from the final value of $\xo\,M_{\rm jet}$
in Fig.~\ref{fmjet} and that quoted by Caratti o Garatti et
al.~(\cite{cagana}) ($M_{\rm acc}\simeq3.4\times10^{-3}$~\Msun).
One obtains
$R_{\rm e}\simeq7.5\times10^{-5}/(3.4\times10^{-3}\,\xo)\simeq2.2\times10^{-2}/\xo$.
Since by definition $\xo\le1$, we conclude that $R_{\rm e}$ must be greater
than a few percent. Vice versa, assuming that at least 10\% of the infalling
material will be redirected into outflow, one has $R_{\rm e}>0.1,$ which sets
an upper limit of $\sim$0.2 on the ionisation fraction, consistent with
the estimate obtained in Paper~I and the values estimated for similar
sources (see Fedriani et al.~\cite{fedr19}).

\section{Summary and conclusions}
\label{scon}

As a follow-up to Paper~I, we monitored the radio continuum
emission of the outburst from the massive YSO \S\ over $\sim$13 months at
six wavelengths
with the VLA.
We also imaged the radio jet at 3~mm with ALMA at two
epochs separated by $\sim$27~months, with an angular resolution $\la$0\farcs1.
Our results indicate that after an exponential increase in the radio flux
in all observed bands, the intensity becomes constant or slightly decreasing.
A comparison of the two ALMA maps shows that the radio jet is
expanding both to the NE and SW, although only the NE lobe was present during our
VLA monitoring. The SW lobe appeared between May 2018 and June 2019, namely
at a much later time than the NE lobe. We believe that this ejection event is
related to the same accretion outburst, which occurred in 2015. We speculate that
the delay between the two lobes might be due to a greater density of the
medium facing the SW lobe, which could curb the expansion of the jet on
that side.

From the analysis of the continuum spectra, we infer that two mechanisms
are needed to explain the observed fluxes: free-free emission at
short wavelengths and non-thermal (probably synchrotron) emission at
long wavelengths. We believe that the latter should become dominant at
frequencies $\la$6~GHz. For this reason, we fitted only the data with
$\nu\ge6$~GHz using a slightly modified version of the jet model adopted
in Paper~I, which works for any opening angle of the jet $<$90\degr. We
conclude that the spectra can be satisfactorily reproduced with an expanding,
decelerating, mono-polar thermal jet that was actively powered by the
outburst until mid-2017. After this date, no more mass is injected into
the lobes and the inner jet radius expands, which, over the long term, is bound
to give rise to one of the knots that characterise thermal jets from YSOs.

\begin{acknowledgements}
A.C.G. acknowledges from PRIN-MUR 2022 20228JPA3A ``The path to star and
planet formation in the JWST era (PATH)'' and by INAF-GoG 2022 ``NIR-dark
Accretion Outbursts in Massive Young stellar objects (NAOMY)'' and Large
Grant INAF 2022 ``YSOs Outflows, Disks and Accretion: towards a global
framework for the evolution of planet forming systems (YODA)''.
R.F. acknowledges support from the grants Juan de la Cierva JC2021-046802-I,
PID2020-114461GB-I00 and CEX2021-001131-S funded by MCIN/AEI/
10.13039/501100011033 and by ``European Union NextGenerationEU/PRTR''.
T.P.R acknowledges support from ERC grant 743029 EASY.
This study is based on observations made under project 16A-424, 16B-427,
and 17B-045 of the VLA of NRAO. The National Radio Astronomy Observatory
is a facility of the National Science Foundation operated under cooperative
agreement by Associated Universities, Inc..
This paper makes also use of the following ALMA data:
ADS/JAO.ALMA\#2018.1.00864.S. ALMA is a partnership of ESO (representing
its member states), NSF (USA) and NINS (Japan), together with NRC (Canada),
NSC and ASIAA (Taiwan), and KASI (Republic of Korea), in cooperation with
the Republic of Chile. The Joint ALMA Observatory is operated by ESO,
AUI/NRAO and NAOJ.
\end{acknowledgements}

\begin{appendix}

\section{Jet expansion law}
\label{aexp}

As discussed in Sect.~\ref{sexp}, the jet is not expanding
at constant velocity during the period of our monitoring, but it appears
to slow down. It is thus necessary to adopt an expression for the maximum
radius, $\rmax$, that properly takes this effect into account.  A reasonable
scenario may be that of a jet confined in a solid angle $\Oj$, with initial
mass $\Mj_0$, which expands with initial velocity $\vj_0$ through a medium
with density $\rho\propto r^{-2}$, where $r$ is the distance from the jet
origin. Because of momentum conservation one can write
\begin{eqnarray}
 \Mj_0 \vj_0 & = & \Mj(t) \frac{\d\rmax}{\d t}  \nonumber \\
  & = & \left[\Mj_0 + \int_{\rmo}^{\rmax} \rho_0\left(\frac{\rmo}{r}\right)^2 \Oj r^2 \, \d{r} \right] \frac{\d\rmax}{\d t}  \nonumber \\
  & = & \left[\Mj_0 + \Oj \rho_0 \rmo^2 \left(\rmax-\rmo\right) \right] \frac{\d\rmax}{\d t}
,\end{eqnarray}
with $\rho_0$ density at radius $\rmo$. The solution of this differential equation is
\begin{eqnarray}
 \Mj_0 \vj_0 t & = & \Mj_0 \left(\rmax-\rmo\right)  \nonumber \\
 & & +~ \Oj \rho_0 \rmo^2 \left[\frac{\rmax^2-\rmo^2}{2} - \rmo\left(\rmax-\rmo\right)\right]
,\end{eqnarray}
which after some algebra gives
\begin{equation}
 \rmax(t) = \rmo + 2 T \vj_0 \left( \sqrt{1+\frac{t}{T}} -1 \right)
  \label{earm}
,\end{equation}
where we have defined $T \equiv \Mj_0/(2\rho_0\Oj \rmo^2 \vj_0)$.

While this expression provides us with a more realistic description of the
jet expansion than the constant velocity assumption adopted in Paper~I, we
stress that it is not to be taken as the real equation of motion of the jet
but as the simplest way to parametrise the observed deceleration of it.

\section{Description of the model}
\label{amodel}

In Paper~I we adopted the jet model by Reynolds~(\cite{reyn})
to describe the integrated flux density of the radio jet from \S. More
specifically, we assumed what is defined as the `standard spherical'
case in Reynolds' Table~1, namely a jet where the opening angle ($\tho$),
internal velocity ($\vo$), ionisation degree ($\xo$), and temperature
($\To$) do not depend on the distance $r$ from the star. Conservation of
mass along the flow implies that the gas number density can be expressed as
$n=\no(r/\ro)^{-2}$.

Despite its simplicity, the model was successful in fitting the observed
spectra in Paper~I. However, Reynolds' equations have been derived under the
assumptions of small $\tho$, an approximation that
is not satisfied by the best fit to the radio spectra obtained in Paper~I
(see Table~3 there), which requires angles as large as $\sim$50\degr. In
order to overcome this limitation, we propose here a slightly modified
version of Reynolds' model that works for any $\tho<90\degr$.

\begin{figure}
\centering
\resizebox{8.5cm}{!}{\includegraphics[angle=0]{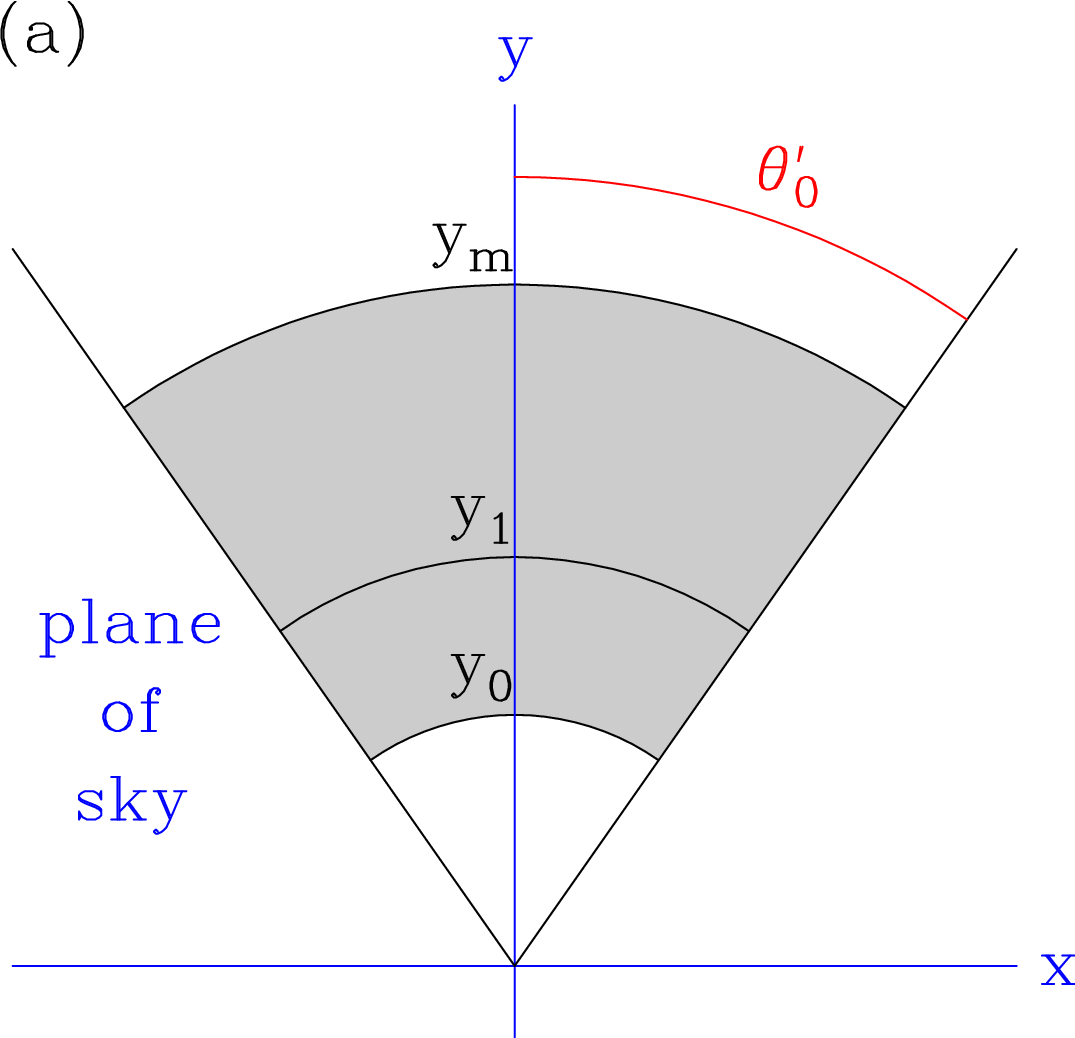}} \\[3mm]
\resizebox{8.5cm}{!}{\includegraphics[angle=0]{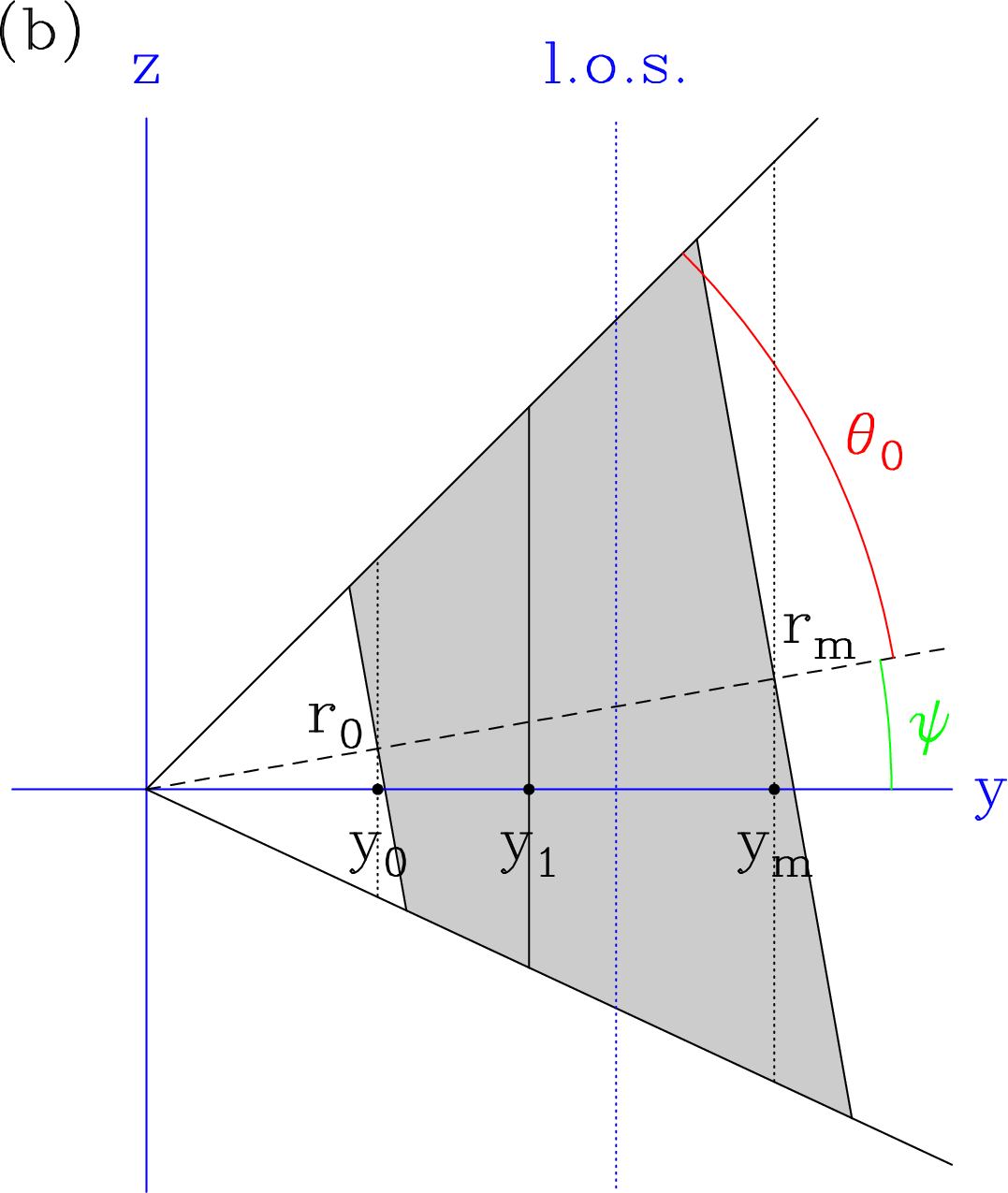}}
\caption{
Sketch of the jet model. The $x$- and $y$-axes lie in the plane of the
sky, and the $z$-axis is parallel to the line of sight. The $y$-axis
coincides with the projection of the jet axis on the plane of the sky. The
dotted line labelled `l.o.s.' denotes a generic line of sight. The star
powering the jet lies at the origin of the coordinate system.
}
\label{fask}
\end{figure}

To allow for an analytic solution of the equations, we maintain Reynolds'
assumption that the opacity depends only on $r$. This is equivalent
to assuming that the jet is not conical but has a pyramidal shape with
two faces parallel to the line of sight. We also assumed that the jet
is delimited by two cylindrical surfaces, with radii $\ro$ and $\rmax$, and that
its axis is inclined by a small angle, $\psi$, with respect to the plane
of the sky. Figure~\ref{fask} schematically illustrates the geometry of the
jet, where the star lies at the origin of the axes and the line of sight
is parallel to the $z$-axis.

Following Reynolds, the absorption coefficient of the
ionised gas and the opacity can be written as
\begin{equation}
 \kappa(r) = \kappa_0 \left(\frac{r}{\ro}\right)^{-4}
,\end{equation}
where
\begin{equation}
 \kappa_0 = a_\kappa \no^2 \xo^2 \To^{-1.35} \nu^{-2.1}   \label{eakz}
,\end{equation}
with $a_\kappa=0.212$ in CGS units, and
\begin{eqnarray}
 \tau(R) & = & \int_{-R\tan(\tho-\psi)}^{R\tan(\tho+\psi)} \kappa \, \d{z} ~=~ \kappa_0 \ro^4 \int_{-R\tan(\tho-\psi)}^{R\tan(\tho+\psi)} \left(R^2+z^2\right)^{-2} \, \d{z} \nonumber \\
         & = & \frac{\kappa_0\ro^4}{2 R^3} \left[ \frac{\tan(\tho+\psi)}{1+\tan^2(\tho+\psi)} + \frac{\tan(\tho-\psi)}{1+\tan^2(\tho-\psi)} + 2\tho \right] \nonumber \\
         & = & \frac{\kappa_0\ro^4}{2 R^3} \left[ \sin(2\tho) \cos(2\psi) + 2\tho \right] \nonumber \\
         & = & \tau_0 \left(\frac{\ro}{R}\right)^3.   \label{eatau}
\end{eqnarray}
Here we have defined the two quantities
\begin{equation}
 R=\sqrt{x^2+y^2}
\end{equation}
and
\begin{equation}
\tau_0 = \kappa_0 \ro \frac{\sin(2\tho) \cos(2\psi) + 2\tho}{2}.
 \label{eatauo}
\end{equation}

To ease the comparison with Reynolds' expressions, we indicate with
$\yo$ and $\ym$ the projections of $\ro$ and $\rmax$ on the plane of the sky,
namely $\yo=\ro\cos\psi$ and $\ym=\rmax\cos\psi$,
and with $y_1$ the value of $R$ at which $\tau(R)=1$.
From Eq.~(\ref{eatau}) one has
\begin{equation}
  y_1 = \ro \tau_0^\frac{1}{3}.    \label{eayu}
\end{equation}

For the calculation of the total flux density of the jet we follow Reynolds'
approach and assume that the emission between $R=0$ and $R=y_1$ is optically
thick, and that between $R=y_1$ and $R=\ym$ is optically thin. Under this
approximation the flux can be written as
\begin{eqnarray}
 S_\nu & = & \frac{1}{d^2} \int_{\yo}^{\ym} B_\nu(\To) \left(1-\e^{-\tau}\right) 2\thop R \, \d{R} \nonumber \\
       & \simeq & \frac{2\thop B_\nu(\To)}{d^2} \left[ \int_{\yo}^{y_1} R \, \d{R} + \int_{y_1}^{\ym} \tau(R) R \, \d{R} \right] \nonumber \\
       & = & \frac{2\thop B_\nu(\To)}{d^2} \left[ \frac{y_1^2-\yo^2}{2} + y_1^3 \left(\frac{1}{y_1}-\frac{1}{\ym}\right) \right]
       \label{easnu}
,\end{eqnarray}
where $B_\nu$ is the Planck function and $\thop$ is the projection of $\tho$
on the plane of the sky.  The two angles are related by the expression
\begin{equation}
 \tan\thop = \frac{\tan\tho}{\cos\psi}.
\end{equation}
We note that it is not strictly correct to integrate from $\yo$ to $\ym$
because the projections of the two circles with radii $\ro$ and $\rmax$
on the plane of the sky are ellipses, and the integral should be made
not only in the variable $R$ but also in the azimuthal angle. However,
the approximation adopted by us is acceptable for small $\psi$.

The expression of $S_\nu$ in the case of a jet totally thin ($y_1<\yo$)
or thick ($y_1>\ym$) can be obtained in a similar way, by considering
only the relevant approximation in the argument of the integral in
Eq.~(\ref{easnu}). In conclusion, one obtains
\begin{eqnarray}
S_\nu & = & \frac{2\thop B_\nu(\To)}{d^2} \nonumber \\
& & \times \left\{ 
\begin{array}{lcl}
 y_1^3 \left(\frac{1}{\yo}-\frac{1}{\ym}\right) & \Leftrightarrow & y_1\le\yo \\
\left[ \frac{y_1^2-\yo^2}{2} + y_1^3 \left(\frac{1}{y_1}-\frac{1}{\ym}\right) \right]
            & \Leftrightarrow & \yo<y_1<\ym \\ 
 \frac{\ym^2-\yo^2}{2} & \Leftrightarrow & y_1\ge\ym
\label{eaflux}
\end{array}
\right. 
.\end{eqnarray}
We note that this expression gives the total flux emitted by a single jet
lobe, whereas Eq.~(5) of Paper~I takes into account both lobes. The
different approach is justified by the fact that our new findings have
proved that only the NE lobe was present during our monitoring (see Sect.~\ref{smmem}).

The quantity $y_1$ is a function of $\no$, in addition to other parameters. However, following Reynolds, it may be convenient to express it in terms of the mass
loss rate of the jet, $\dot{M}$, which is obtained by integrating
the flux of mass through the inner surface of the jet. Obviously, $\dot{M}$
does not depend on the inclination of the jet with respect to the line of sight,
so we can simplify the calculation by assuming $\psi=0$. Hence, we obtain
\begin{eqnarray}
 \dot{M} & = & \int_{-\ro\tan\tho}^{\ro\tan\tho} \mu \no \frac{\ro^2}{\ro^2+z^2} \vo \frac{\ro}{\sqrt{\ro^2+z^2}} 2\tho\ro \, \d{z} \nonumber \\
         & = & 4 \mu \no \vo \tho \ro^4 \int_{0}^{\ro\tan\tho} \left(\ro^2+z^2\right)^{-\frac{3}{2}} \, \d{z} \nonumber \\
         & = & 4 \mu \no \vo \tho \ro^2 \frac{\tan\tho}{\sqrt{1+\tan^2\tho}}
         \label{eamdot}
,\end{eqnarray}
where $\mu$ is the mean particle mass per hydrogen atom (we assume
$\mu=1.67\times10^{-24}$~g) and $\vo \frac{\ro}{\sqrt{\ro^2+z^2}}$ is
the component of the velocity perpendicular to the inner surface of the
jet. From this it is trivial to express $\no$ as a function of $\dot{M}$,
and from Eqs.~(\ref{eakz}), (\ref{eatauo}), and~(\ref{eayu}) one obtains
\begin{equation}
 y_1 = \left( \frac{a_\kappa}{16\mu^2} \MM^2 \To^{-1.35} \nu^{-2.1} \,
              \frac{1+\tan^2\tho}{\tan^2\tho} \,
              \frac{\sin(2\tho)\cos(2\psi)+2\tho}{2\tho^2} \right)^\frac{1}{3}
,\end{equation}
where we have defined $\MM\equiv\xo\dot{M}/\vo$.

Finally, a quantity of interest for our purposes is the ionised mass of
the jet, which is given by the expression
\begin{eqnarray}
 \Mi & = & \int_{-\tho}^{\tho} \d{\theta} \int_{\ro}^{\rmax} R \, \d{R} \int_{-R\tan\tho}^{R\tan\tho} \mu \xo \no \frac{\ro^2}{R^2+z^2} \, \d{z} \\
  & = & 4 \mu \xo \no \tho^2 \ro^2 \left(\rmax-\ro\right) \nonumber \\
  & = & \MM \tho \frac{\sqrt{1+\tan^2\tho}}{\tan\tho} \left(\rmax-\ro\right)
  \label{eamjet}
,\end{eqnarray}
where we have used Eq.~(\ref{eamdot}) to replace $\no$.

\end{appendix}

\end{document}